\documentclass[prb,showpacs,showkeys,secnumarabic]{revtex4}
\usepackage{amssymb}
\usepackage{graphicx}
\usepackage{subfigure}
\usepackage{float}
\usepackage{subeqnarray}
\usepackage{color}
\usepackage{bm}
\usepackage{amsmath}

\begin{document}

\title{Electron Transmission Across Normal Metal-Strained Graphene-Normal Metal Junctions}
\author{Weixian Yan$^1$\footnote{The corresponding author. Tel: +86 13734017012, E-mail address: wxyansxu@gmail.com (W.-X. Yan).},  Guo Min$^2$}
\affiliation{$^1$College of Physics and electronics, Shanxi University, 030006, People's Republic of China}
\affiliation{$^2$Institute of Theoretical Physics, Shanxi University, 030006, People's Republic of China}


\begin{abstract}
  The transmission of the electron across the single normal metal-graphene (NG) and normal-metal-graphene-normal-metal (NGN) junctions has been investigated. For the single NG junction, the profile of the maximum transmission which has been plotted against the dimensionless interface hopping respectively bears similarity to that of the conductance of the system. The minor effect of the incidence energy on transmission can also be found in conductance of the single NG junction whose tunneling behavior poses a striking difference from that of the NGN junction. Concerning with NGN junction, the transmission and conductance show more abundant structures when subjected to different incidence energies, interface hopping, and strain strengths. The increase of strain strength always induces more resonance peaks at different angles in transmission and can therefore enhance the conductance. The increase of length of the middle graphene segment can accommodate more quasi-resonance states, leading to the more resonance peaks and richer structures in transmission.  In both single NG and NGN junctions, the increase of the wavefunction period on metal side(s) can be observed due to the enhancement of strain strength, which can serve as the sensor for the detection of the strain strength in graphene.
\end{abstract}


\maketitle

\section{Introduction}
In recent decades, the miraculous discovery of the single layer of carbon atoms, graphene\cite{first,neto}, has stirred up enthusiasm 
on the development of various types of the two-dimensional materials\cite{chiu,nakh}, anticipating to revolutionize the next-generation electronic 
and optoelectronic applications and bridge the gaps of different disciplines of physical sciences\cite{neto,caldas,bauer}.  Many kinds of two-dimensional materials with different physical mechanisms have emerged, to name a few, transition metal dichalcogenides\cite{kolobov}, silicene\cite{takeda,aufray}, molecular Dirac system\cite{kajita}, etc. where novel electronic and optical properties, size effect, discrete symmtries,  and gauge symmetry may be conceived.

As the promising substitutes for semiconductor heterostructures, the exploration on the salient electronic, transport and optical properties of graphene heterostructures\cite{allain,sinha,barbier,nasci,biswas,barbier2,martino,garg,kkim,dellanna2,azadi,zeb,hanggi,martino3,orozco,yan,arovas,fuchs,kaur,chant,liu,champo,xiecqu,zhaobnu} becomes urgent and essential. In recent decades, the theoretical and experimental studies on graphene quantum wells (including p-n junctions)\cite{barbier,nasci,azadi,arovas}, superlattices\cite{barbier2,martino,martino3} and  graphene quantum dots\cite{hewag,schulz,schnez} have been carried out by many groups.  Most of the studies focused on the graphene-graphene heterostructures, whose different segments are tuned by the different  magnetic fields  (vector potentials)\cite{martino,martino3,fuchs,biswas} and  different electric potentials to create wells and barriers\cite{martino,barbier}, 
 while the  confinement potentials for graphene quantum dots are vector potentials\cite{yanmpb} as well as static and time-dependent electrical potentials\cite{schulz,orozco} in different circular regions;  as for hybrid graphene heterostructures, there are graphene-metal  heterostructures\cite{giovannetti,cusati,blanter,schomerus,chaves,ysang} and  graphene-superconductor heterostructures where the proximity of graphene to superconducting layer result in the specular Andreev reflections\cite{beenakker,maiti,goudarzi}.

As far as the most easily accessible measurement on electrical transport of two-dimensional layered materials is concerned, the contact between graphene and metal is the most interesting and important\cite{giovannetti,cusati,blanter,schomerus} since the two-dimensional graphene hybrid electronic devices inevitably have metal electrodes. The metal electrode also can tune the energy band structure \cite{cusati} and electron transport properties of graphene\cite{giovannetti,blanter,schomerus}, and the contact resistance between the metal-graphene interface is an important indicator that affects the performance of graphene devices. The metal contact provides the electron reservoir\cite{giovannetti,blanter} 
 and the major source of scattering around the boundary of the metal-graphene interface\cite{blanter,schomerus,giovannetti,sang}, therefore the study on scattering around the metal-graphene 
hetero-structure has become a necessity. To this end, Blanter and Martin studied the transport properties of electrons at the NG and NGN junctions 
 in the framework of tight-binding approach; the reflection/transmission and conductance of both NG contact and NGN contacts were analyzed analytically for propagating and evanescent modes in the small energy regime\cite{blanter}.
 Inpsired by their work,  we have incorporated the strain\cite{naumis,champo} into the graphene segment and studied the effect of the strain on the scattering at the 
 NG ang NGN junctions. The analytical expression for transmission across the NG and NGN junctions 
 has been obtained, and the numerical computation on transmission has been carried out by focusing on the (maximum) transmission and conductance against the 
 strain strengths, dimensionless interface hopping and incidence energies/angles. Furthermore, the wavefunction on each segment of NG and NGN have been presented 
 under the different strains. The work is organized as follows, the model of the NG and NGN junctions incorporating the strain is 
presented in Section II; the  numerical analysis is presented in section III,  and Section IV concludes our work.

\section{The model analysis of the strained NG and NGN junctions as well as the derivation of transmission}
\subsection{The model analysis and incoporating strain into graphene}
  In the framework of the tight-binding approximation, the Hamiltonian governing the motion of the electron in pristine graphene can be written as:
$
\widehat{{\cal H}}=\sum\limits_{{\bf R},n}\left[t_{\bm{R},n} a^\dag_{{\bf R}}b_{{\bf R}+\delta_n}+h.c.\right],
$
where $a_{{\bf R}},b_{{\bf R}+\delta_i}$ stand for the annihilation operators of the electron at position ${\bf R}$ (atom $A$) and ${{\bf R}+\bm{\delta}_i},$
 (nearest-neighboring atoms $B$) with $ \bm{\delta }_1 =\frac{a_g }{2} (1,\sqrt{3} )^T$, 
 $\bm{\delta}_2 =a_g (-1,0)^T$,  $\bm{\delta }_{3} =\frac{a_g}{2} (1,-\sqrt{3})^T$. When the strain is applied, the  Hamiltonian governing the electrons in strained graphene can be rewritten as follows,
\begin{eqnarray}
H=-\sum\limits_{\bm{R}^{\prime},n}t_{\bm{R}^{\prime},n} a_{\bm{R}^{\prime}}^\dag b_{\bm{R}^{\prime} +{\bm {\delta}}_{n}^{\prime}}  +h.c
\end{eqnarray}
where $\bm{R}^{\prime}$ is the new position vector for atom $A$, and $\bm{R}^{\prime} +{\bm {\delta}}_{n}^{\prime}$ are the new position vectors for the nearest neighbor atoms $B$.
 The relationship between ${\bm {\delta}}_{n}^{\prime}$ and ${\bm {\delta}}_{n}$ obeys the transformation rules:
~${\bm{\delta'}_n}=({\bf I}+{\bm \varepsilon}){\bm{\delta}_n}$, where ${\bm \varepsilon}$ is the strain matrix and can be expressed as\cite{pereira},
\begin{eqnarray} &&
[\bm{\varepsilon}]=\xi\begin{pmatrix}   \cos^2\theta-\sigma \sin^2\theta & (1+\sigma) \sin\theta \cos\theta \\
(1+\sigma) \sin\theta \cos\theta  &   \sin^2\theta-\sigma \cos^2\theta \end{pmatrix}
\end{eqnarray}
where $\sigma$ is the Posisson ratio and equals to $\sigma \simeq 0.165$ and $\xi$
 represents the strain strength. When the strain

\begin{figure}
\centering 
\subfigure[]{ 
\label{Fig01a} 
\includegraphics[height=5cm,width=0.375\textwidth]{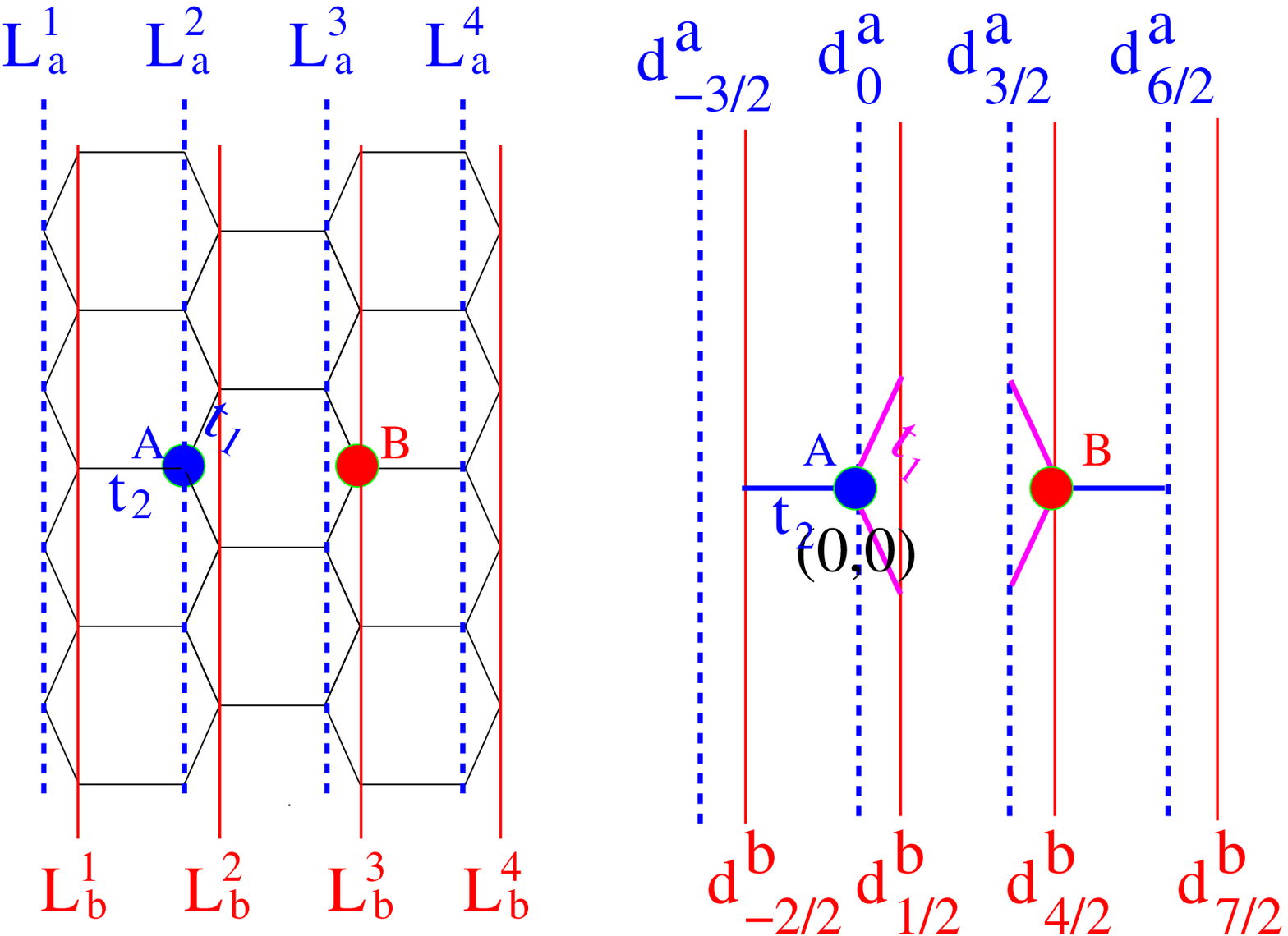}} 
\subfigure[]{ 
\label{Fig01b} 
\includegraphics[height=5cm,width=0.375\textwidth]{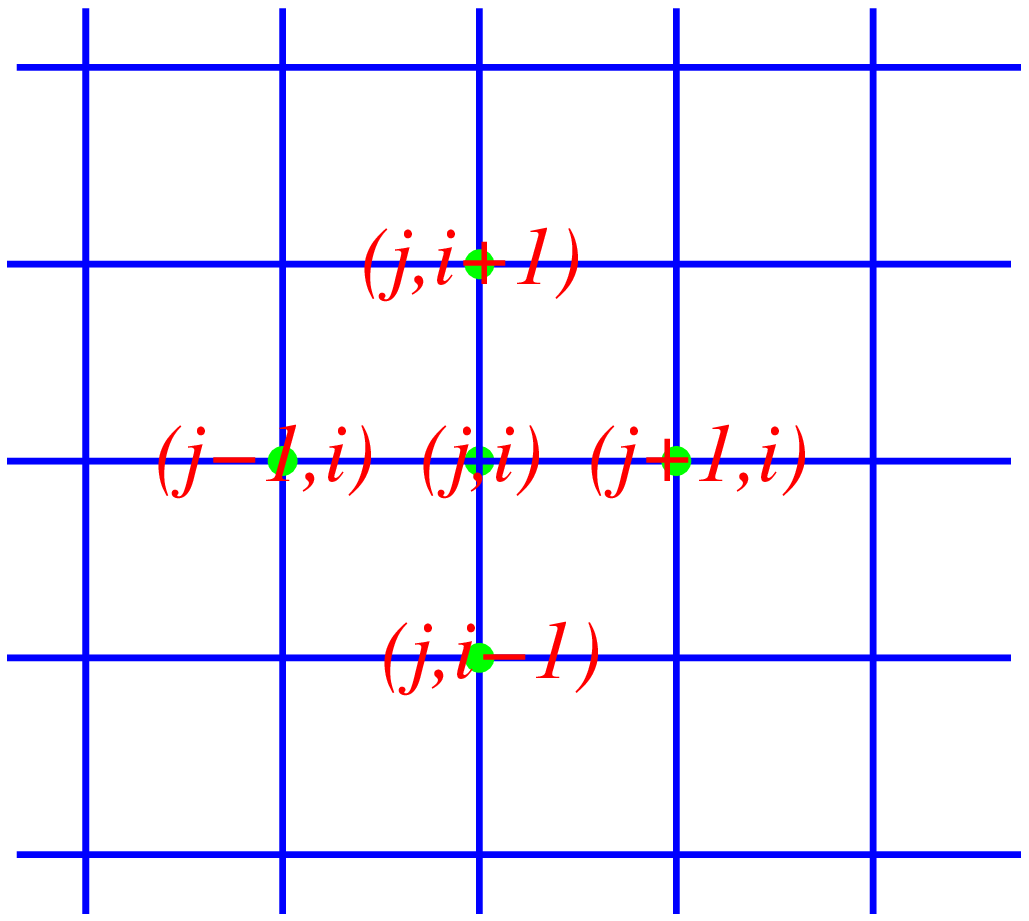}}
\centering 
\subfigure[]{ 
\label{Fig01c} 
\includegraphics[width=0.75\textwidth]{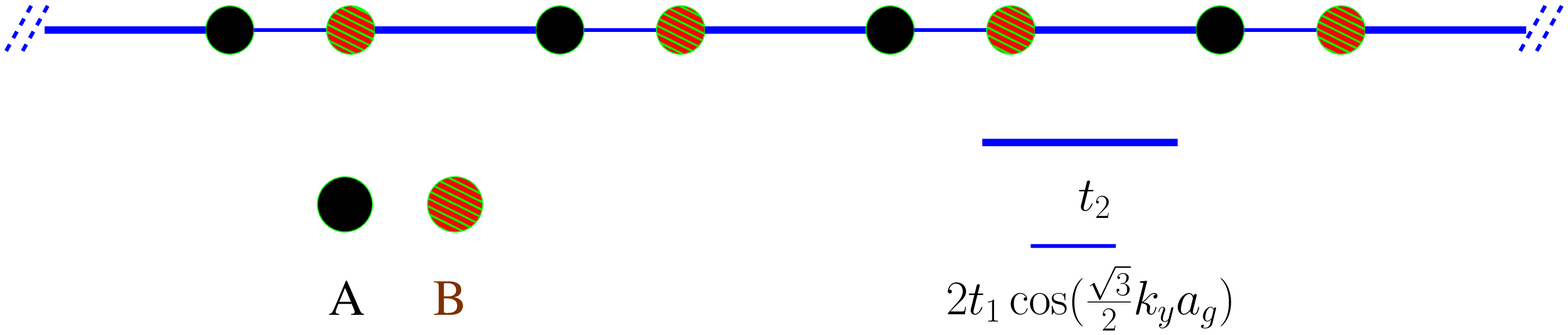}} 
\caption{ (a) Left: Part of strained graphene where the blue dashed line $L_a^i$ (i=1,2,3,4) represents the locations of $A$ atoms; and 
 red solid line  $L_b^i$ (i=1,2,3,4) represents the locations of $B$ atoms.
 Right: the blue dashed lines and red solid lines are the same as those on the left, while $L^i_{a(b)}$ have been replaced by lattice wavefunctions $d^a_{j}$ where 
 $j=\cdots -3/2,0,3/2,\cdots$ for $A$ atoms, and $j=\cdots -2/2,1/2,4/2,\cdots$ for $B$ atoms. (b) A part of 2-D metal square lattice. (c) The 1-D  effective chain for strained graphene with the conserved transverse momentum $\hat{p}_y$. } 
\end{figure}

 is applied along the armchair direction ($\theta=0^\circ$ or $X$-direction) in the 
graphene segment, the nearest neighbor vectors $\bm{\delta}_i^{\prime}$  under strain 
can  be calculated as, $\bm{\delta }_1^{\prime} \approx  \left( \left(\frac{a_g }{2} + u_2 \right),\frac{\sqrt{3} }{2} a_g \right)^T,
\bm{\delta }_2 ^{\prime} =  \left( \left (-a_g -u_1 \right), {0} \right)^T,
\bm{\delta }_{3} ^{\prime} \approx \left( \left(\frac{a_g }{2} + u_2 \right),-\frac{\sqrt{3} }{2} a_g \right)^T$.
where  $\sigma \xi $ has been plausibly neglected based on the fact that $\sigma =0.165$ and maximum strain strength $\xi$ is assumed to be 0.2 in our work, 
$u_1=a_g \xi, u_2=a_g\xi/2$, superscript $T$ stands for 
 transposition of the matrix. The so-called Dirac point can be obtained by minimizing the following altered eigen-energies,
\begin{eqnarray}
 E'({\bf k})=\pm {|\sum\limits_i t_i e^{i{\bf k}\cdot {\bm \delta}'_i}|} =\pm \left|t_2 +t_1 e^{-i\bm{k}\cdot \bm{a}^\prime_1 } +t_1 e^{-i\bm{k}\cdot \bm{a}^\prime_2 } \right|, 
\end{eqnarray}
where the hopping integral $t_i$ can be
  estimated by a empirical relation $t_i=t_0 e^{-3.37(|{\bm \delta}'_i|/a-1)}$
   due to alteration of the nearest-neighbor vectors\cite{sood1,wong}, and the relation of $t_3=t_1$ has been used.
By minimizing the above new eigenenergy with respect the $k_x$, and $k_y$ respectively $\frac{\partial E}{\partial k_{x} } =0 $, $\frac{\partial E}{\partial k_y } =0$, 
 the new Dirac point can be obtained as, $\pm {\bm {K}}=\left(0,\pm \frac{2\sqrt{3}}{3a_g} \arccos \left(-\frac{t_2 }{2t_1 } \right)\right)^T\equiv\pm (K_x,K_y)$.
In the low energy regime, the eigen-energy of the strained graphen can be obtained as $E=\pm \sqrt{\frac{3}{4} a_g^2 q_{y}^2 \left(4t_1^2 -t_2^2 \right)+\frac{9}{4} a_g^2 t_2^2 q_{x}^2 }
$ where $q_x, q_y $ are the crystal momentum along $x,y$ directions respectively.
  By expanding the energy $E'({\bf k})$ around  ${\bm {K}}$, the effective Hamiltonian  can be obtained as follows:
   \begin{eqnarray} 
	\label{pycomh2}
	&&
H_{\rm {eff}}=\hbar\begin{pmatrix}   0 & v_x q_x - i v_y q_y \\
v_x q_x + i v_y q_y  &   0 \end{pmatrix}~,
\end{eqnarray}
where $v_x=\frac{\sqrt{3}}{2}a(1+ {\cal S})\sqrt{4t^2_1-t^2_2},  v_y=\frac{3}{2}a(1- \sigma{\cal S})t_2\approx v_F$, 
and $v_F$ is the so-called Fermi velocity in pristine graphene.

From the above analysis, it is self-evident that the strain 
 exerted along the armchair direction produces negligible effect on the the graphene lattice along the $Y-$direction ($\sigma\xi\approx 0$).
 Therefore, it is plausible to assume $[\hat{p}_y,\hat{H}]=0$ in the the strained graphene.
As shown in the left panel of Fig.1(a), assume that the pseudospin spinor corresponding to the sublattices $A,B$ of the graphene at position $\bm {r} $ can be represented as:
  $\psi_g(\bm{r})=(d^{a} (\bm{r}), d^{b}\bm{r})^T$, we can derive the effective Hamiltonian for strained graphene as follows (for detail, see Appendix),
\begin{eqnarray}
~ H^{\textrm{eff}}_{\textrm{SG}}=&&-\sum\limits_i \bigg[\left(2t_1 \cos\left(\frac{\sqrt{3}}{2}k_ya_g\right) |'3i/2'\rangle \langle [3i/2+1/2]| + t_2 |'3i/2'\rangle \langle [3i/2-1]|\right) \nonumber \\ &&  +h.c.\bigg]
\end{eqnarray} 
	where X-coordinates for A-atoms and B-atoms are respectively represented by index of single quotes $('i')$ and square bracket $[i]$ instead of the real X-cordinates as illustrated by the right panel of Fig.1(a).
	Substitute the expansion $\psi=\sum\limits_j d_j^a |'j'\rangle + \sum\limits_m d_m^b|[m]\rangle$ into Schr\"odinger equation, 
 then the equation governing the strained graphene can be obtained as follows,
\begin{eqnarray}
\label{ver1_7}
&&
Ed_{3i/2}^{a}=-t_2 d_{3i/2-1}^{b}-2t_1\cos\frac{\sqrt{3}k_ya_g}{2} d_{3i/2+1/2}^b    \nonumber \\ &&
Ed_{3i/2+1/2}^{b}=-t_2 d_{3i/2+3/2}^{a}-2t_1\cos\frac{\sqrt{3}k_ya_g}{2} d_{3i/2}^a~,
\end{eqnarray}                            
which are graphically illustrated by Fig.1(c).

For NG junction where the graphene side is subject to the uniform strain along the armchair direction as shown in Fig.2(a) 
where the metal is arranged on the left side, while the strained graphene is on the right side. 
The metal has been assumed to be a square tight-binding lattices which can be described by the Hamiltonian\cite{gpzhang}: 
 ${\cal H}_s=-t_s\sum\limits_{i,j} \left[\left(C_{i,j} C^\dag_{i+1,j}+C_{i,j} C^\dag_{i,j+1}\right)+h.c\right]$,  where  $C_{i,j}$ ($C_{i,j}^\dag$) operators represents the electron 
 annihilation (creation) operator on site $(i,j)$ in the square tight-binding lattices. For metal side, the wavefunction along the $Y-$direction  can be expressed by a plane wave, i.e., $\left| ij \right\rangle =\left| i \right\rangle e^{ik_y a_s j} $
where  $a_s $ is the lattice constant of the square lattice. Then the effective site-energy for this two-dimensional square 
lattice can be computed as $2t_s\cos k_y a_s $, and the effective hopping remains to be $t_s$. The equation governing this two dimensional perfect square tight-binding lattice can be derived with help of the effective Hamiltonian $H^{\textrm{eff}}_{\textrm{NM}}$ defined in Appendix, 
\begin{eqnarray}
\label{ver1_8}
(E+2t_s \cos k_y a_s)c_j=-t_s(c_{j+1}+c_{j-1})
\end{eqnarray}
where the effective site energy $2t_s \cos k_ya_s$ can be clearly seen.

\begin{figure}
\centering 
\subfigure[]{ 
\label{Fig02a} 
\includegraphics[width=0.6\textwidth]{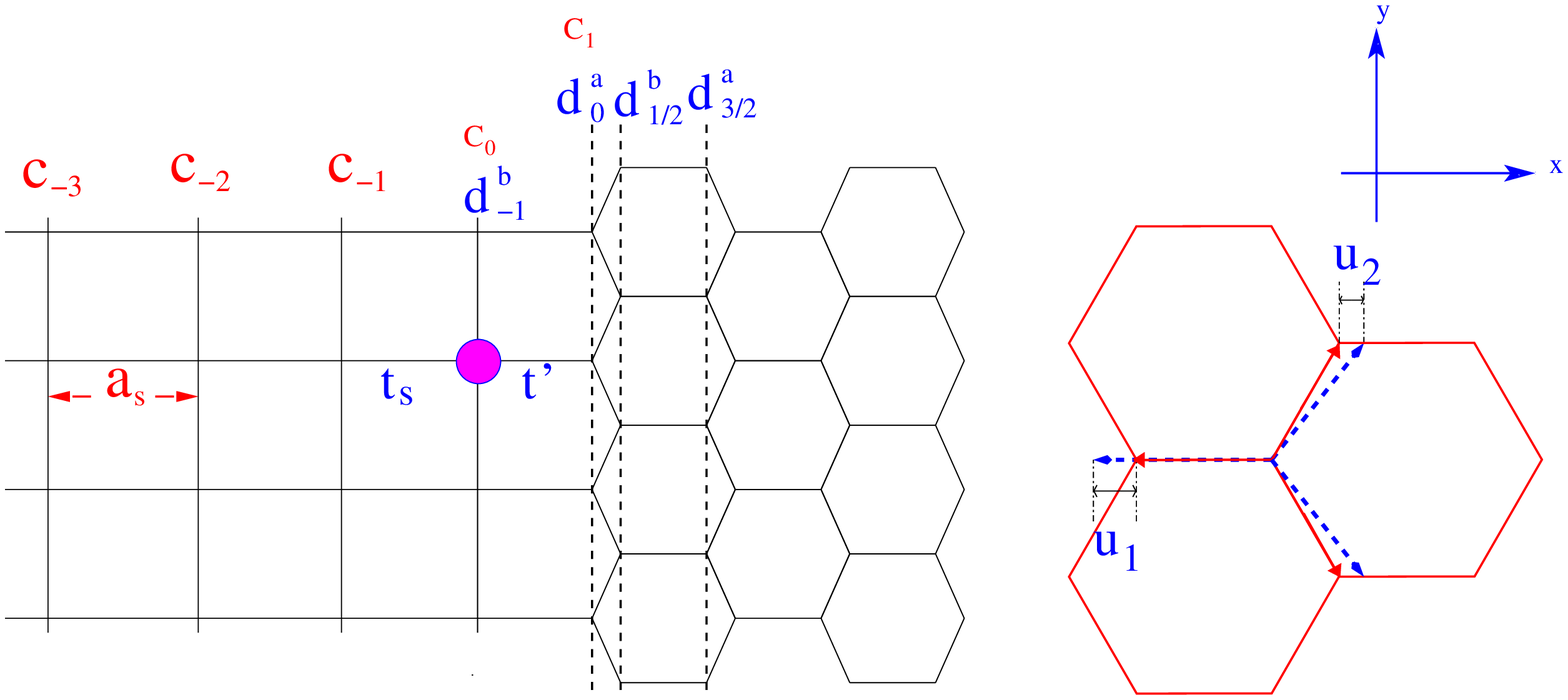}} 
\subfigure[]{ 
\label{Fig02b} 
\includegraphics[width=0.8\textwidth]{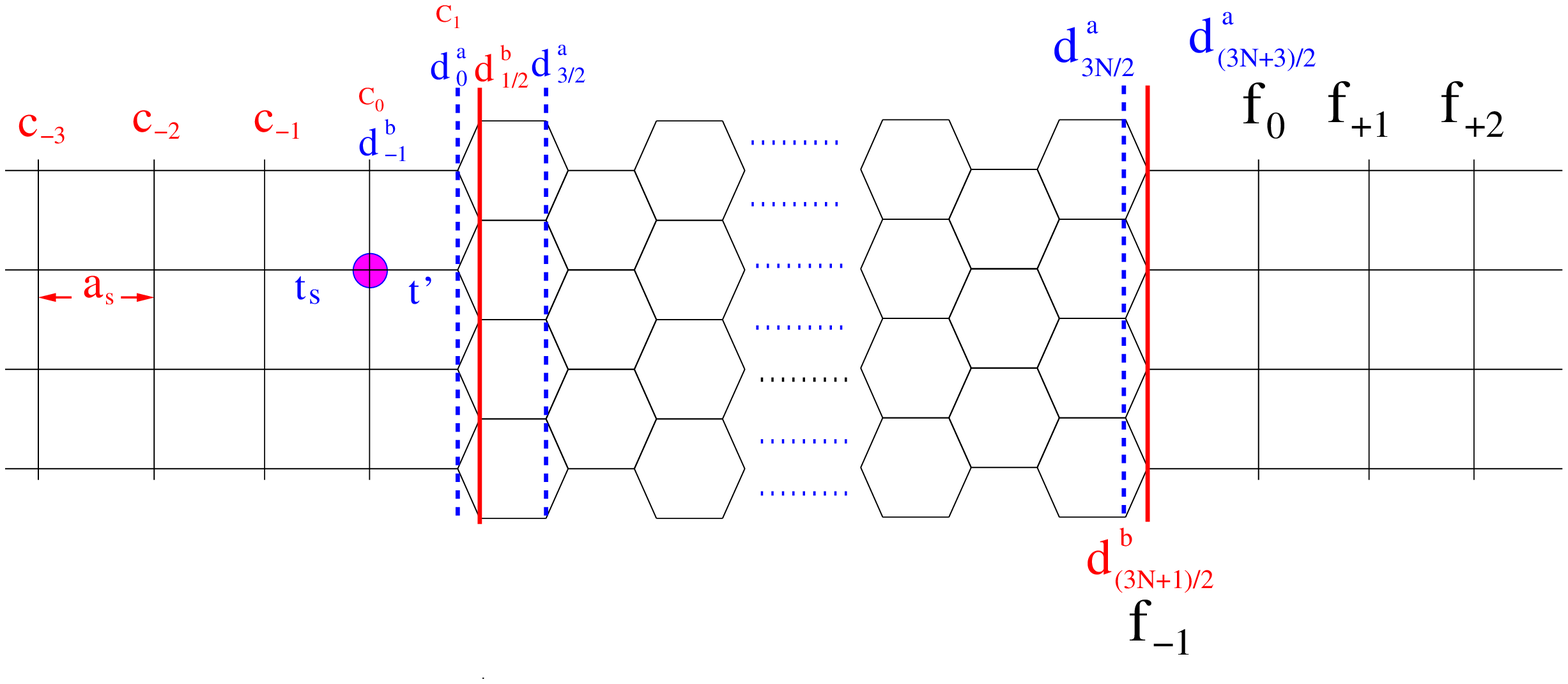}}
\caption{ (a) Left side: the metal-graphene (NG) junction where the coefficients $c_i$ and $d_j$ represent the on-site wavefunctions for metal and strained graphene respectively;
Right side: three red solid  vectors are nearest-neighbor vector basis $\bm{\delta}_i, (i=1,2,3)$ for pristine graphene, while three blue dashed vectors are 
nearest-neighbor vector basis $\bm{\delta^\prime}_i, (i=1,2,3)$ for strained graphene, $u_1$ and $u_2$ are the length changes along $x$ direction. 
(b) The metal-graphene-metal (NGN) junctions with $f_i$ being the on-site outgoing wavefunction of the right-side metal, $N$ (odd integer) stands for the number of the periods of graphene. } 
\label{Fig.lable1} 
\end{figure}

	For the single NG junction, the relation between the wavefunctions at the interface of the heterostructure of metal and strained graphene can be 
	 established by employing the following equation (see Appendix for derivation),  
\begin{eqnarray}
\label{ver1_9}
&& Ec_{0} =-2t_{s} c_{0} \cos k_y a_s -t_{s} c_{-1} -t^{\prime} d_{0}^{a}   \nonumber \\ &&
 Ed_{0}^{a} =-t' c_{0} -2t_1 \cos \frac{\sqrt{3} k_y a_g }{2} d_{1/2}^{b} \nonumber \\ &&
Ed_{3i/2}^{a}=-t_2 d_{3i/2-1}^{b}-2t_1\cos\frac{\sqrt{3}k_ya_g}{2} d_{3i/2+1/2}^b ~~(i=0,1,2,\cdots)\nonumber \\ &&
Ed_{3i/2+1/2}^{b}=-t_2 d_{3i/2+3/2}^{a}-2t_1\cos\frac{\sqrt{3}k_ya_g}{2} d_{3i/2}^a~~(i=0,1,2,\cdots)
\end{eqnarray}
where  $t_s$ stands for  hopping in the metal and $t_1,~t_2$ stand for hoppings in the strained graphene, 
and $t^\prime $ stands for the hopping between  graphene and metal at the heterostructure interface, 
and the strain effect on the hopping integrals has been taken into account.

\subsection {Derivation of transmission for NG and NGN junctions}
When there is an incoming plane wave $e^{ik_s x}e^{i q_s y}$ ($q_s$ is the momentum along $Y$ direction) on the metal side and propagates to the graphene side and reflects at the metal-graphene interface, 
 and then the wave function at position ${\bm {r}}$ on the metal lattices can be expressed
	as $c(\bm {r})=\left(e^{ik_s x}+r e^{-ik_s x}\right)e^{i q_s y}$.
By taking account of the outgoing wave in the strained graphene side and the strain-induced variation of the nearest neighbor vectors $u_1$, and $u_2$ along the $X$-direction, which are defined previously and schematically illustrated on the right panel of Fig.2(a), the following equation can be obtained,
\begin{eqnarray}
Ed_{0}^{a} =-\left[t_2 e^{-iq_{x} \left(a_g +u_1 \right)} +2\left(t_1 \cos \frac{\sqrt{3} k_y a_g }{2} e^{iq_{x} \left(\frac{a_g }{2} +u_2 \right)} \right)\right]d_{0}^{b} 
\equiv \lambda^r_{ab}d_{0}^{b} 
\end{eqnarray}
where $q_x=q\cos\theta$, $q_y=q\sin\theta$, and $q=\sqrt{q^2_x+q^2_y}$.
The reflectance coefficient can be determined by employing Eqs.(7)-(9),  and its final form can be expressed as follows,
\begin{eqnarray}
\label{ngexact}
r=\frac{E\exp(ik_s a_s)-\beta\lambda^r_{ab}\exp(iq_x(a_g+u_1))}{E\exp(-i k_s a_s)-\beta \lambda^r_{ab}\exp(iq_x(a_g+u_1))}
\end{eqnarray}  
where $\beta=\frac{t'^2}{t_s t_2}$. In order to takes into account of overall strain-induced effect on the graphene hopping integral term, the dimensionless parameter $\beta_0$ defined for the normal-metal-pristine graphene in Ref.(\onlinecite{blanter}) has been employed as one of the main parameters to replace the dimensionless parameter $\beta=\frac{{t^\prime}^2}{t_s t_2}=\beta_0\cdot t_g/t_2$  ($t^\prime\sim 0 $ indicates the perfect reflection of particles at the normal metal-graphene interface). 
                                 
Concerning with NGN junction as shown in Fig.2(b), the normal metal on the left and right sides can be 
 deemed as the the left and right electrode respectively. The wavefunctions on the left electrode and right electrode can be respectively expressed 
as\cite{blanter}: $c(r)=\left( \exp(ik_s x+ r \exp(-ik_s x))\right) \exp(ik_y y)$, and $f(r)=\omega \exp(i(k_s[x-L])) \exp(ik_y y)$.
In the middle graphene segment, there are left and right-moving waves at each lattices $(\bm r)$
and the corresponding lattice wavefunctions are denoted as $d^{a(b)}_\ell(\bm r)$, $d^{a(b)}_r(\bm r)$, for detailed meaning of symbols, see Fig.2(b).
With help of same method as that used to derive the equations for NG junction, the following equations on the 
 coefficients of $d^a_\ell(0), d^a_r(0), r, \omega$ can be derived as, 
\begin{eqnarray}
\label{ker}
~~&& t_s \left(1+r\right)=t^\prime \left(d^a_\ell(0)+d^a_r(0)\right) \nonumber \\ &&
+E\left( \frac{1}{\lambda^\ell_{ab}}e^{+iq_{x}a_g \left(1+\xi\right)}d^a_\ell(0) +  \frac{1}{\lambda^r_{ab}}e^{-iq_{x}a_g \left(1+\xi\right)}d^a_r(0) \right)=\frac{t^\prime}{t_2} \left(\exp(-ik_sa_s)+r\exp(ik_sa_s)\right) \nonumber \\ &&
\omega t^\prime \exp(ik_s a_s)=t_2\left(d^a_\ell\left(0\right)\exp(-i(L+a_g)(1+\xi)q_x) 
+d^a_r\left(0\right)\exp(i(L+a_g)(1+\xi)q_x)\right) \nonumber \\ &&
\omega t_s=t^\prime \left(+\frac{E}{\lambda^\ell_{ab}}d^a_\ell(0)e^{-iL(1+\xi)q_x} 
+\frac{E}{\lambda^r_{ab}}d^a_r(0)e^{iL(1+\xi)q_x} \right)
\end{eqnarray}
where $\lambda^\ell_{ab}=-\left[t_2 e^{+iq_{x} \left(a_g +u_1 \right)} +\left(2t_1 \cos \frac{\sqrt{3} k_y a_g }{2} e^{-iq_{x} \left(\frac{a_g }{2} +u_2 \right)} \right)\right]$.
Solving the above equations, we can obtain, 
\begin{eqnarray}
~\omega = \frac{1}{\left(\sum_i^3 {\cal D}_i\right)}E t^\prime t_s \left(-t^\prime+t_2 e^{2 i a_s k_s}\right) e^{i (\xi +1) a_g q_x-i a_s k_s} \left(\lambda^\ell_{ab} e^{2 i l (\xi +1) q_x}-\lambda_{ab}^r e^{2 i (\xi +1) q_x \left(a_g+l\right)}\right)
\end{eqnarray}
where ${\cal D}_1, {\cal D}_2, {\cal D}_3$ can be expressed as,
\begin{eqnarray}
&& {\cal D}_1=t^\prime t_s \left(E^2 t^\prime+t_2 \lambda^\ell_{ab} \lambda_{ab}^r\right) \left(-1+e^{2 i (\xi +1) q_x \left(L+a_g\right)}\right) e^{i \left((\xi +1) q_x \left(L+a_g\right)+a_s k_s\right)} \nonumber \\ &&
{\cal D}_2=E {t^\prime}^3 e^{2 i a_s k_s} \left(\lambda^\ell_{ab} e^{i (\xi +1) q_x \left(3L+2 a_g\right)}-\lambda_{ab}^r e^{i (\xi +1) q_x \left(L+2 a_g\right)}\right) \nonumber \\ &&
{\cal D}_3=-e {t^\prime}_2 t_s^2 \left(\lambda^\ell_{ab} e^{i  (\xi +1) q_x L}-\lambda_{ab}^r e^{i (\xi +1) q_x \left(3L+4 a_g\right)}\right)~.
\end{eqnarray}

\section{Numerical analysis and discussion}
In this section,  the effects of strain strngth, incidence energy and dimensionless interface hopping $\beta$ on the transmission for NG junction and NGN heterostructure have been analyzed by the numerical analysis. For single NG junction with $\beta_0=0.2$ in Fig.3(a) and Fig.3(b), whether incidence energies are 100 meV or 500 meV, the strain is conductive to the transmission and the maximum transmission $\tau_m$ increases with the increase of the strain, and the transmission is marginally altered by the increase of incidence energy (by comparing Fig.3(a) with Fig.3(b), and Fig.3(c) with Fig.3(d)). Since the conductance  is closely associated with the area  under the transmission curve in the figure, the increase of the strain can clearly enhance the conductance. When $\beta_0$ is increased to 0.8, the increase of the strain suppresses the capability of electron tunneling from metal side to the graphene side in single NG junction as illustrated by panels of Fig.3(c) and Fig.3(d).

\begin{figure}[H] 
\centering 
\label{Fig03} 
\includegraphics[height=8cm,width=0.8\textwidth]{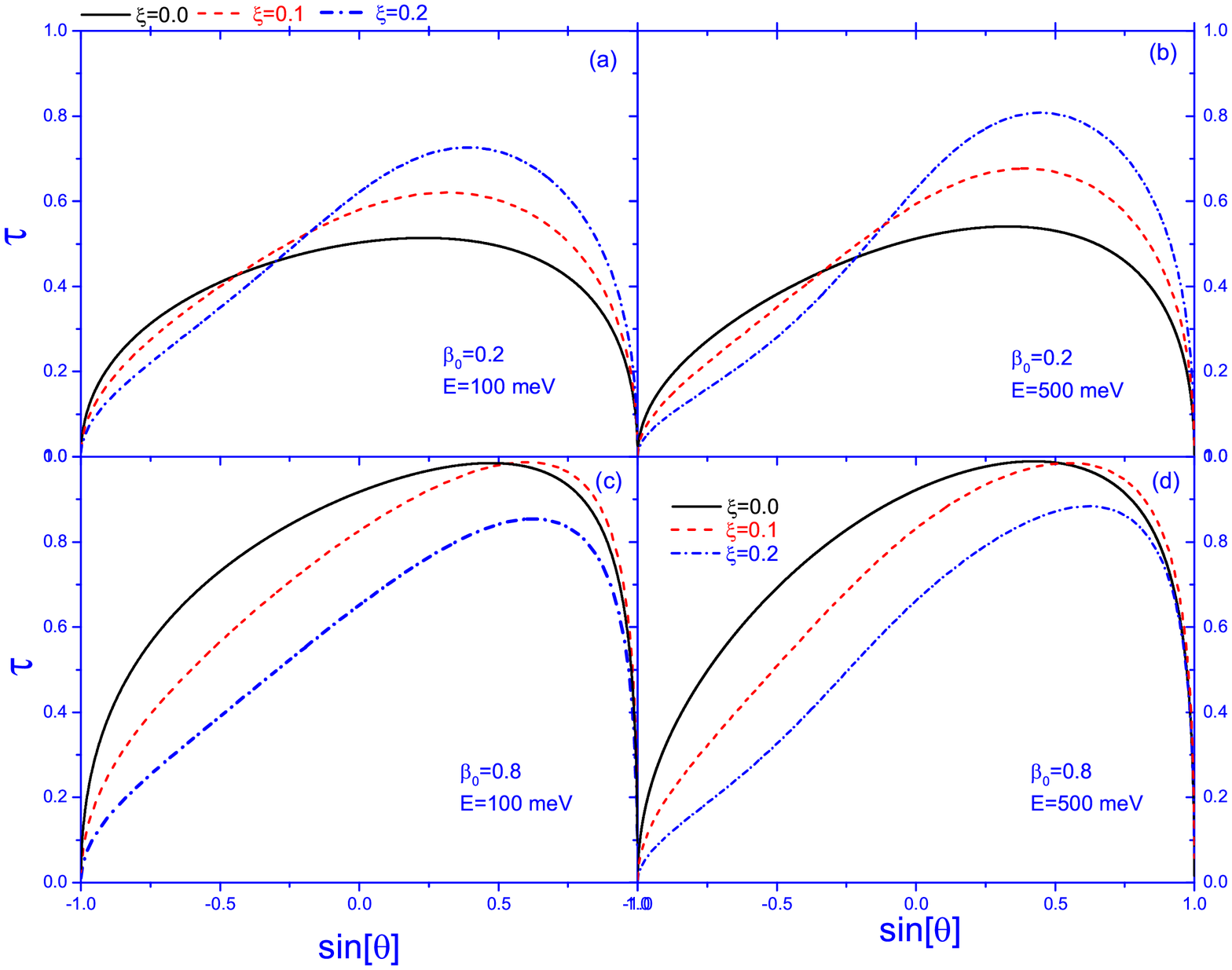}
\caption{The transmission $\tau$ against $\sin[\theta]$ for NG junction under three different cases of strains $\xi$ with incident energies of $E=100$ meV, $500$  meV. (a) $\beta_0=0.2, E=100$ meV,  (b) $\beta_0=0.2, E=500$ meV, (c)  $\beta_0=0.8, E=100$ meV  (d) $\beta_0=0.8, E=500$ meV, the cases of $\xi=0.0, 0.1 $ and $0.2 $ are respectively represented by black solid line, red dashed line, and  blue dash-dot line.} 
\end{figure}

While for the transmission of NGN junctions with periods of $N=80$ as shown in Fig.4, the black solid lines and red dotted lines respectively stand for the exact and approximate results (Eq.(10) of Ref.[\onlinecite{blanter}]) in the strainless $(\xi=0)$ case, red dashed lines and blue dash-dot lines respectively represent the strained cases of $\xi=0.1,0.2$. As shown in the top panel of Fig.4(a), the exact result is consistent with the approximate result in the case of small incidence energy, and two results clearly deviate from each other when the incidence energy has been increased to 500 meV, which is illustrated by black solid lines and red dotted lines in the top panel of Fig.4(b). In contrast to the single NG junction, the transmission in NGN junction undergoes drastic changes due to increases of both the incidence energies and strains. By comparing two panels of Fig.4, it can be seen that at the same incidence energy and $\beta_0$, the increase of the strain can lead to more resonance peaks. When incidence energy is increased to 500 meV in Fig.4(b), much more resonance peaks emerge, showing a striking difference 
 to the single NG junction. In order to further elucidate this  difference, the dimensionless conductance $G/G_0$ $(G_0=2e^3 WV_G/(2\pi)^2 3a_gt_g \hbar)$ has been mumerically computed for single NG and NGN junctions and shown in Fig.5(a) and Fig.5(b) respectively. It is worth noting here that the conductance is closely related  but not proportional to the area under the transmission curve. For single NG junction, Fig.5(a) shows that the increase of incidence energy almost produce negligible effect on the conductance, demonstrating a remarkable distinction from the NGN junction in Fig.5(b) where the conductance depends sensitively on the incidence energy.
 In the NGN junction, the conductance oscillates with the incidence energy and can be enhanced by the presence of strain, more peaks and valleys in conductance emerge with the increase of the strain strength $\xi$.

For single NG junction, the maximum transmission $\tau_m$ and the corresponding angle $\theta_m (\sin[\theta_m])$ against the dimensionless $\beta_0$ under three different cases of strain   strength have been respectively shown in the top and bottom panels in Fig.6.  The transmission across the NG interface is completely blocked for an opaque interface $t^\prime=0$\cite{blanter}, the increase of $\beta_0 $ (or $\beta$) does not necessarily result in the improvement of the transmission. 
 In order to describe the changing tendency,  the optimum transmission $\tau^{opt}_m$ which stands for the peak value of the maximum transmission $\tau_m$ has been introduced. 

\begin{figure}[H] 
\centering 
\subfigure[]{ 
\label{Fig04a} 
\includegraphics[height=7cm,width=0.9\textwidth]{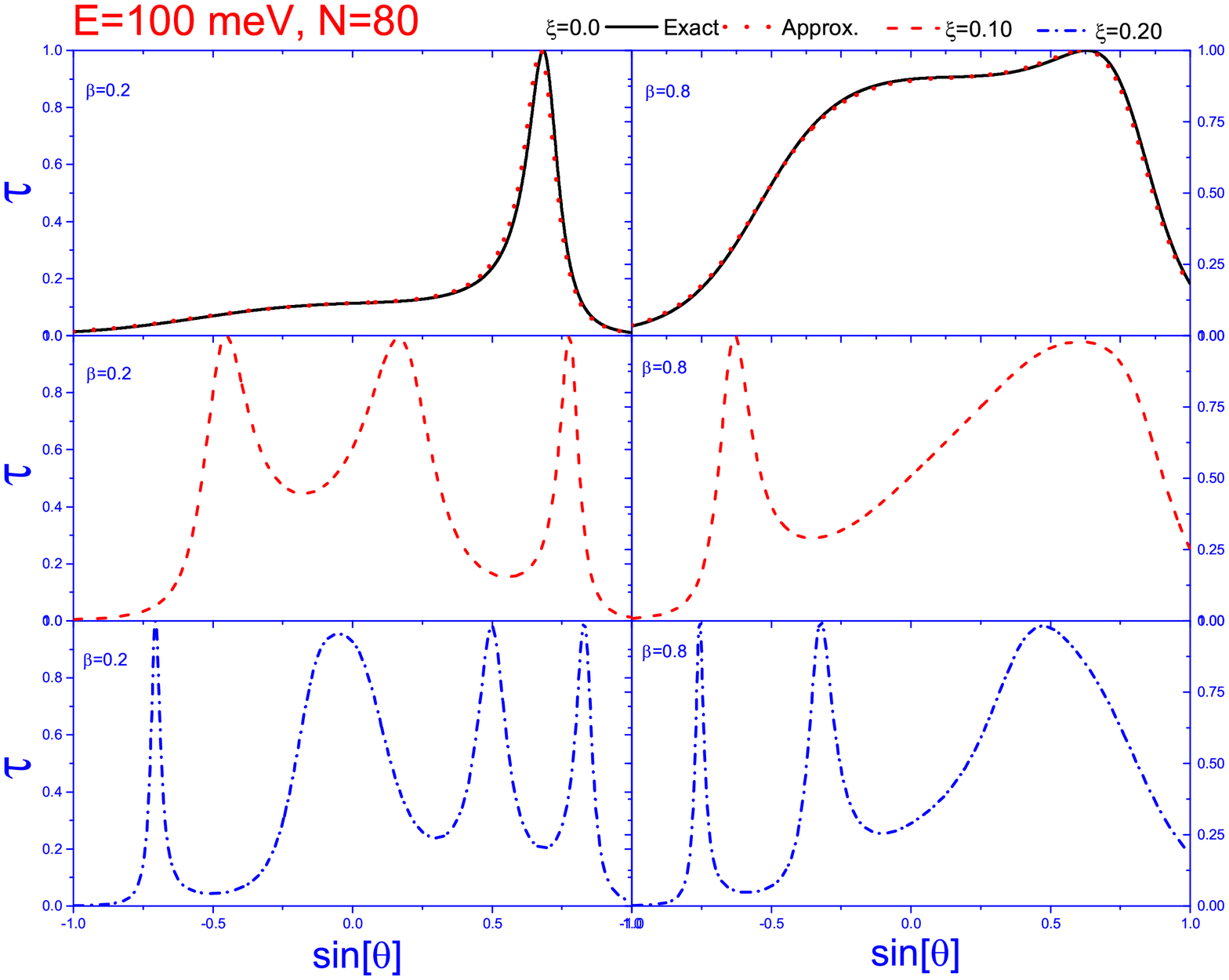}} 
\subfigure[]{ 
\label{Fig04b} 
\includegraphics[height=7cm,width=0.9\textwidth]{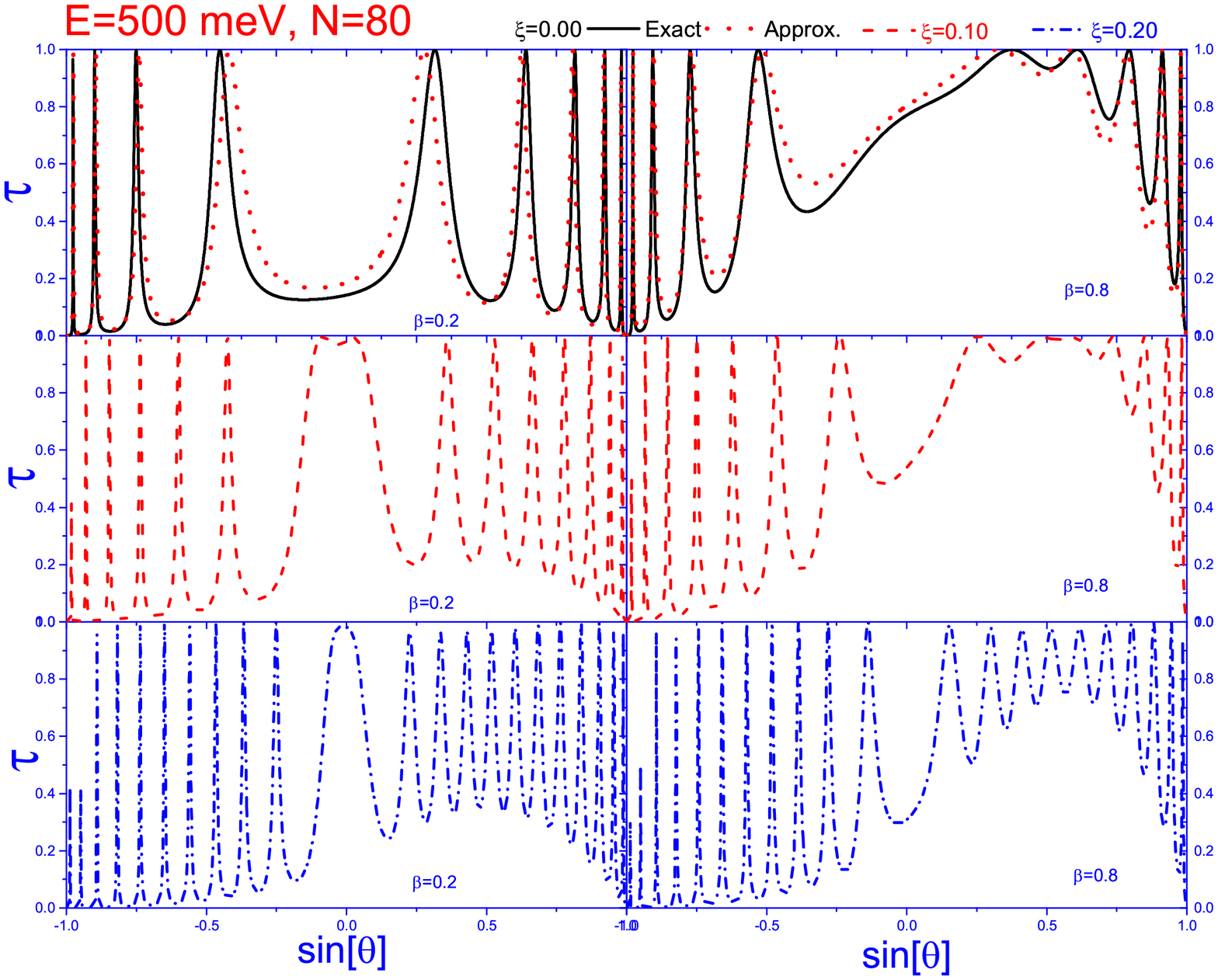}}
\caption {The transmission $\tau$ against $\sin[\theta]$ for NGN junction with periods of $N=80$ under three different cases of strains $\xi$ with incident energies of $E=100$ meV, $500$  meV. (a) $ E=100$ meV,  (b) $ E=500$ meV,  the cases of $\xi=0.0, 0.1 $ and $0.2 $ are respectively represented by black solid line, red dashed line, and  blue dash-dot line, while the approximate result for strainless case of $\xi=0.0$ are represented by red dots.}
\end{figure}

As illustrated  in the figure, the optimum transmission moves to the smaller $\beta_0$  with the increase of the strain, while the optimum angle $\theta^{opt}_m$ increases with the increase of the strain (further away from $\theta=0^\circ$). 
The increase of dimensionless $\beta_0$ beyond the optimum values as indicated by arrows in Fig.6 will rapidly suppress the transmission maximal, which is further aggravated by the presence of the strain.
The corresponding  transmission maximum angle $\theta_m$ draws close to $\theta=0^\circ$ direction and the further increase of the strain shows negligible effect on the transmission maximum angle $\theta_m$, when dimensionless hopping $\beta_0$ attains a greater value.
What is perhaps more interesting is the numerical result  shown in Fig.7 where the conductance of the single NG junction has been plotted  against the dimensionless $\beta_0$.
The conductance can roughly be divided into three regions which are partitioned by two red dotted lines $L_1$ and $L_2$.
When $\beta_0<L_1$, the increase of strain enhances the conductance, when $\beta_0>L_2$,  the increase of strain suppresses the conductance.
In the region within $L_1$ and $L_2$, there is no monotonic relation between the strain and conductance.
  By comparing the curve profiles of top panel of Fig.6 and Fig.7, it is found that the two curves bear similarity to each other, indicating that locating the maximum transmission 
	 $\tau_m$ is an effective and appreciably equivalent way of describing the conductance for the single NG junction. 
	
\begin{figure}[H] 
\centering 
\subfigure[]{ 
\label{Fig05a} 
\includegraphics[height=4cm,width=0.45\textwidth]{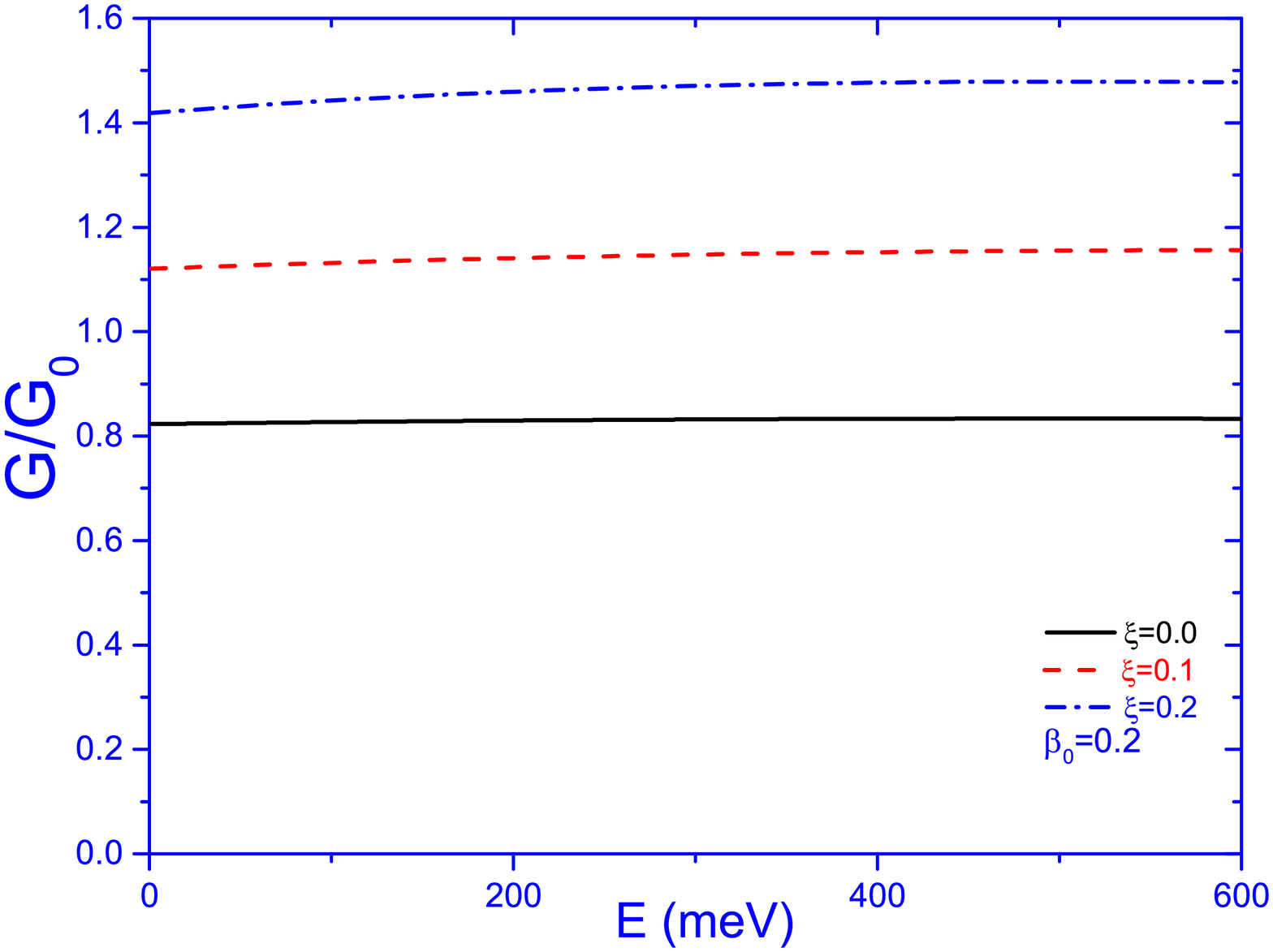}} 
\subfigure[]{ 
\label{Fig05b} 
\includegraphics[height=4cm,width=0.45\textwidth]{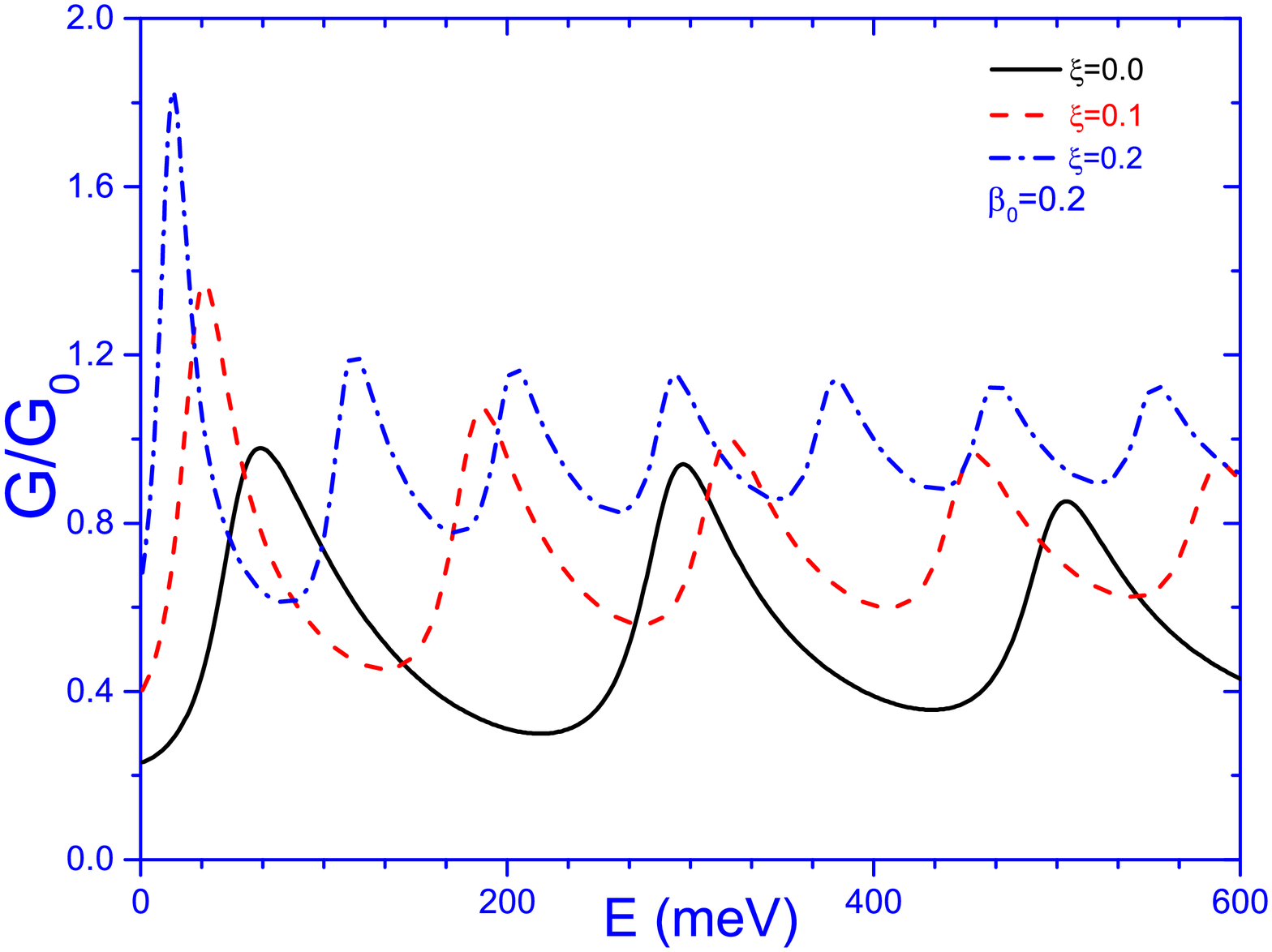}}
\caption {The conductance $G/G_0$ against the incidence energy $E$ for NG junction  and NGN junction, (a) NG junction, (b) NGN junction.}
\end{figure}

\begin{figure}[H] 
\centering 
\label{Fig06} 
\includegraphics[height=6cm,width=0.9\textwidth]{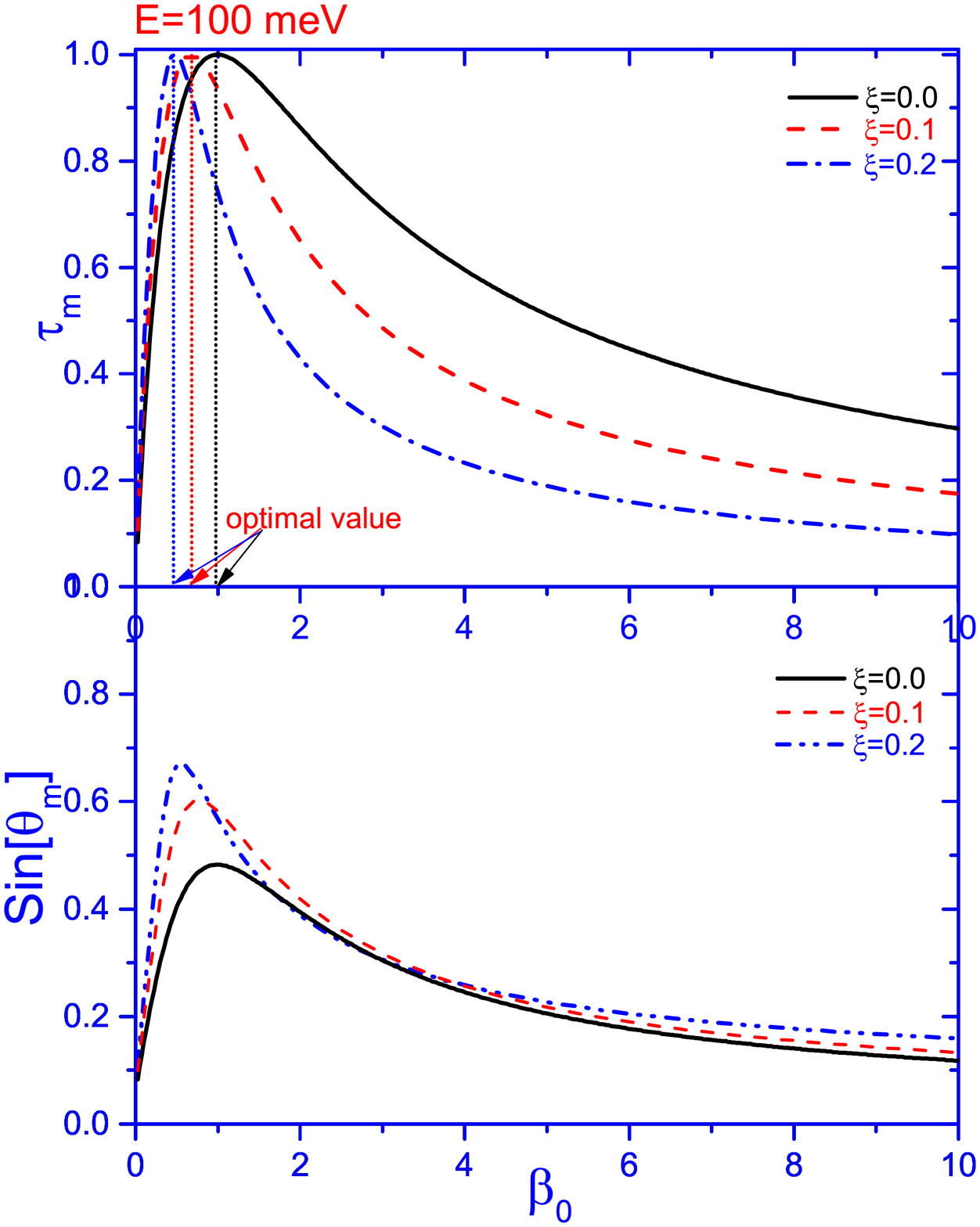}
\caption {The maximum transmission $\tau_m$ and the corresponding angle $\theta_m (\sin[\theta])$ against the dimensionless $\beta_0$ are shown in the upper and lower panels respectively under three cases of strain $\xi=0.0, 0.1, 0.2$ with the incidence energy of $E=100$ meV, the optimum transmission  $\tau^{opt}_m$ refers to the peak value of the $\tau_m$.
}
\end{figure}

\begin{figure}[H] 
\centering 
\label{Fig07} 
\includegraphics[height=4cm,width=0.9\textwidth]{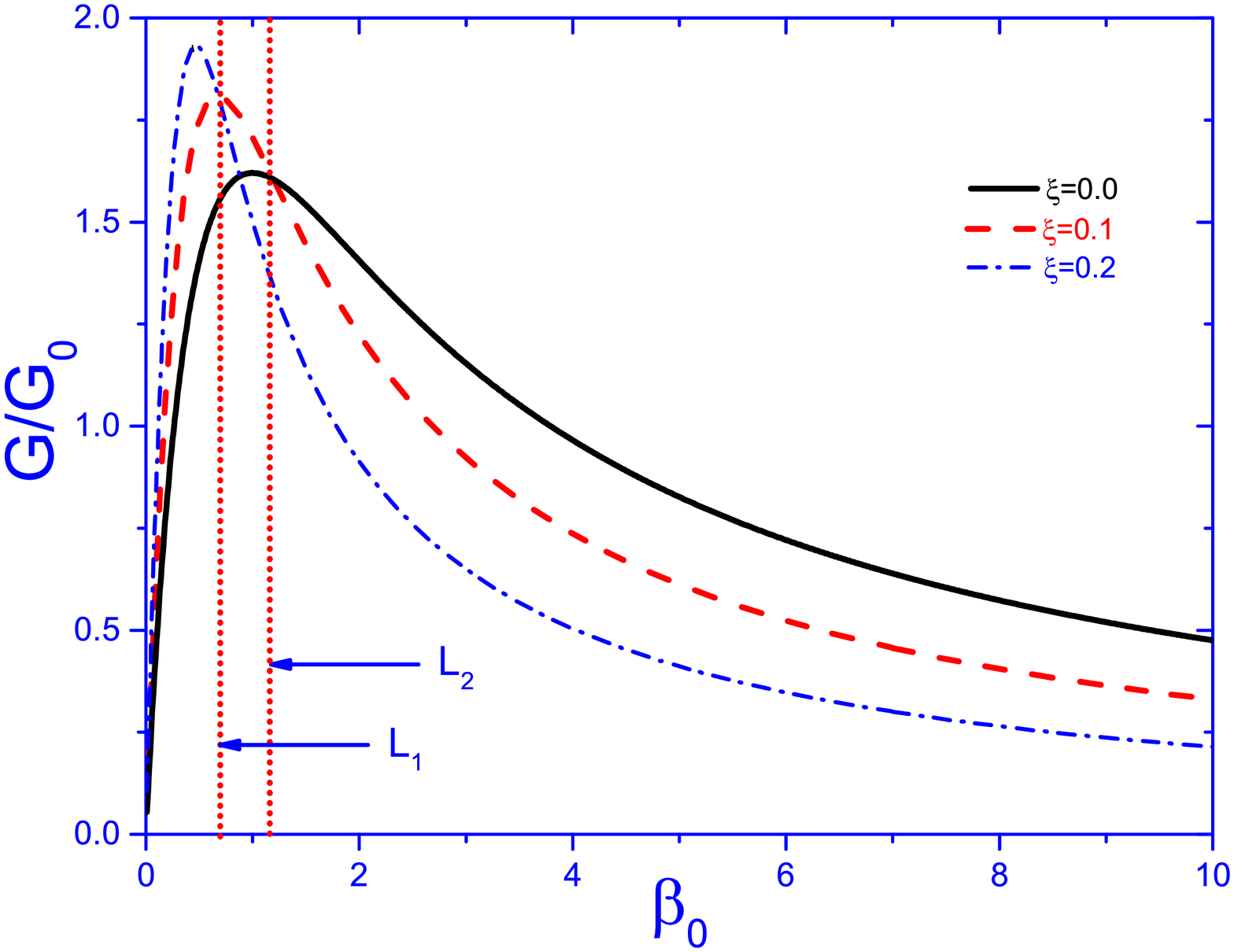}
\caption {The maximum transmission $\tau_m$ and the corresponding angle $\theta_m (\sin[\theta])$ against the dimensionless $\beta_0$ are shown in the upper and lower panels respectively under three cases of strain $\xi=0.0, 0.1, 0.2$ with the incidence energy of $E=100$ meV, the optimum transmission  $\tau^{opt}_m$ refers to the peak value of the $\tau_m$.
}
\end{figure}

As illustrated in the above numerical analysis for single NG junction, the transmission sensitively depends on strain strength $\xi$ and dimensionless interface hopping $\beta_0$. In order to give a panoramic view of  how the two parameters influence the transmission, the contour plot 
 of the transmission against both $\xi$ and $\beta_0$ has been presented in Fig.5.
 The incidence energies in the upper and lower two panels  are respectively $E=100$ meV,  and $E=500$ meV, while the right two panels depicts the maximum transmission 
 $\tau_m$, and the left two panels illustrates the the angles $\sin \theta_m$ corresponding to the maximum transmission $\tau_m$.
 In all these panels, the larger maximum transmission/angles (red-color) and the smaller maximum transmission/angles (blue color) are separated by a white curved strip. 
 As indicated in the two right panels, the increase of the incidence energies produces negligible effect on the maximum transmission.
 When incidence energy is smaller, say $E=100$ meV, the  larger maximum transmission angles mainly distributed in the small range of dimensionless interface hopping (see upper left panel), while the increase of the incidence energies will make the  larger maximum transmission angles distribute within the larger range of the dimensionless interface hopping as illustrated in the lower left panel.
To put it another way, in the case of the larger incidence energy, the increase of the strain drives the larger maximum transmission to distribute within the greater range of dimensionless interface hopping and 
 make the maximum angle deviate further away from the $\theta=0^\circ$. 

\begin{figure}[H] 
\centering 
\label{Fig08} 
\includegraphics[height=6cm,width=0.6\textwidth]{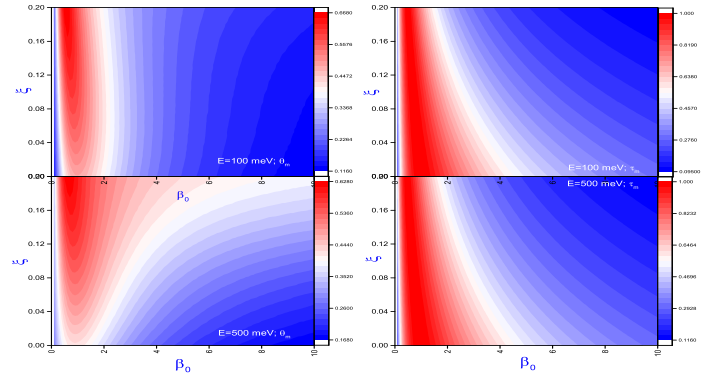}
\caption {The contour plot transmission maximum $\tau_m$ and the corresponding angle $\theta_m$ are plotted against the strain $\xi$ and interface hopping $\beta_0$, the upper two panels and lower two panels 
 are for $E=100$ meV and 500 meV respectively; the  two panels on the left hand and two panels on the right hand are for maximum angle and maximum transmission respectively.
}
\end{figure}

\begin{figure}[H]
\centering 
\subfigure[]{ 
\label{Fig09a} 
\includegraphics[height=4.5cm,width=0.6\textwidth]{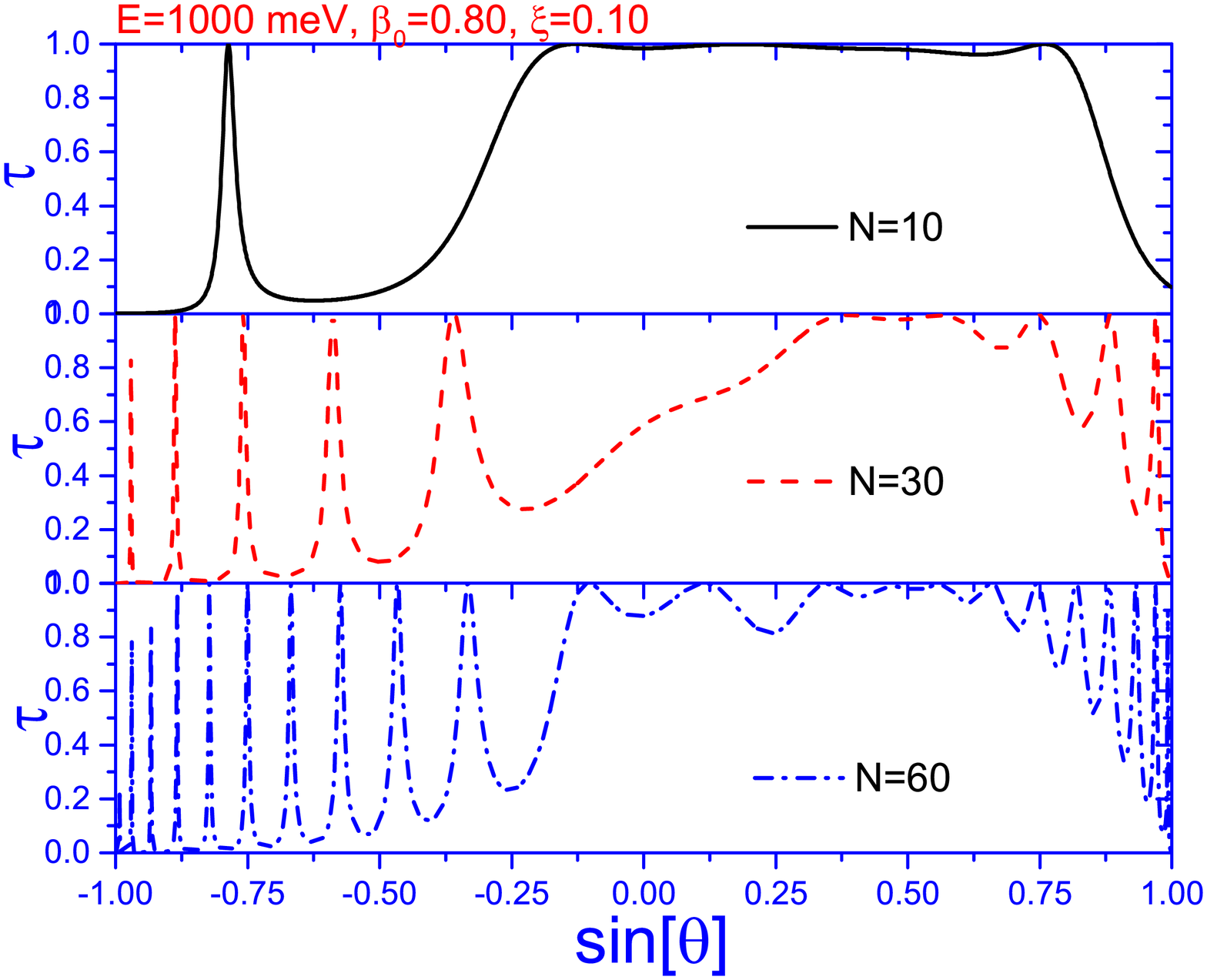}} 
\subfigure[]{ 
\label{Fig09b} 
\includegraphics[height=4.5cm,width=0.6\textwidth]{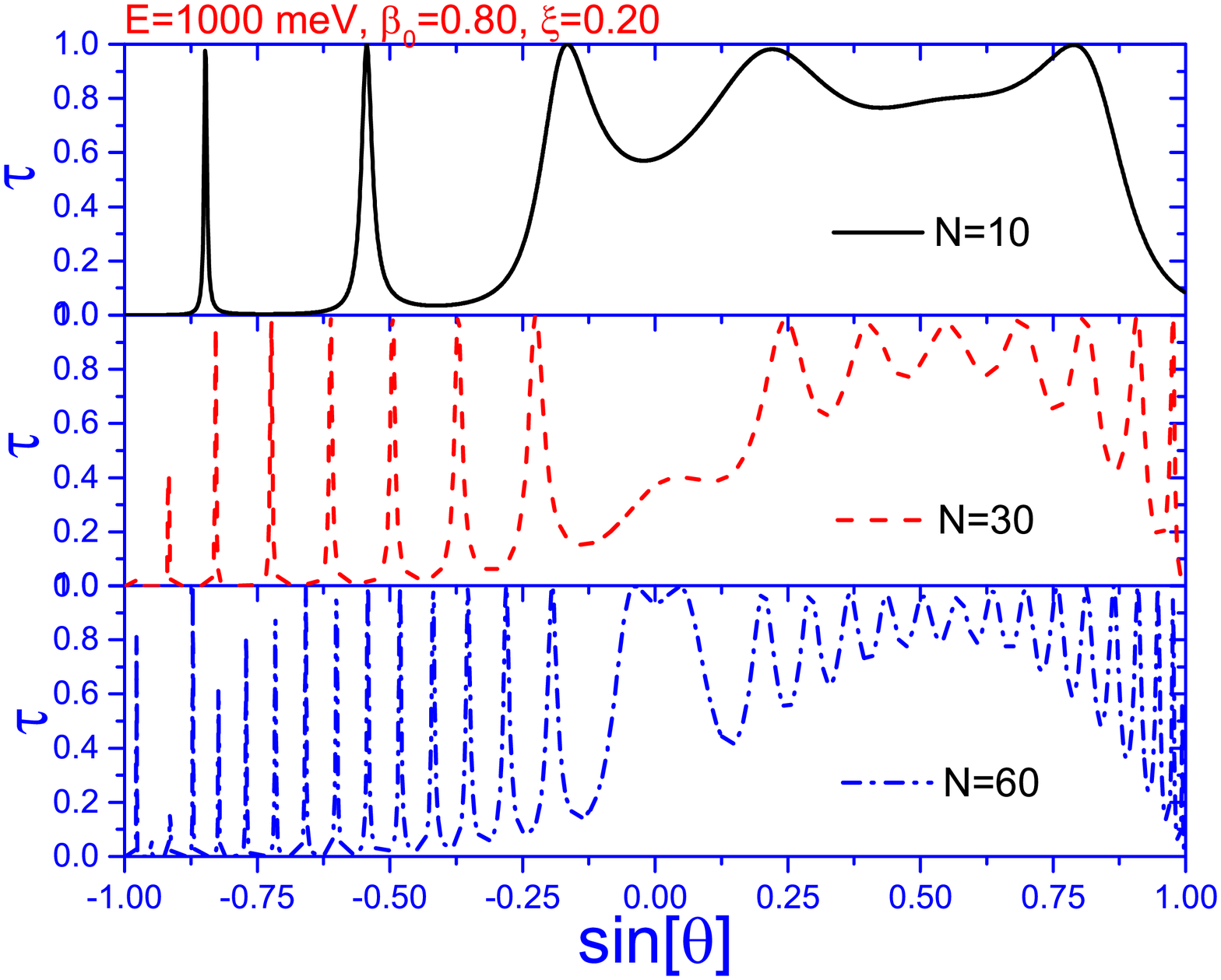}}
\caption{ Transmission $\tau$ against $\sin[\theta]$ under three different cases of lattice periods of graphene (length) $N=10,30, 60$. (a) E=1000 meV, $\beta_0=0.8$, $\xi=0.1$,
(b) E=1000 meV, $\beta_0=0.8$, $\xi=0.2$
}
\label{Fig.lable1} 
\end{figure}

The results for single NG junction presented here have demonstrated that the strain and dimensionless interface hopping play an important role in the 
 electron transmission across the NG interface, while the incidence energy $E$ has a negligible effect on the transmission across the 
 interface, which is in contrast to graphene-graphene heterostructures\cite{barbier}.
 The reason behind such negligible effect of the incidence energy $E$ on the transmission is attributed to the limited match 
 of the longitudinal ($Y$-direction) momentum between the two sides of the interface\cite{blanter} as well as lack of resonance mechanisms which prevail in NGN junction.

In order to investigate how the graphene segment length between the two normal metal sides influence the transmission at constant incidence energy $E (1000$ meV),  dimensionless $\beta_0 (0.8)$.  The strain strengths are  chosen to be $\xi=0.1$ and 0.2 in Fig.9(a) and Fig.9(b) respectively.
With the increase of the number of graphene lattice periods from $N=10$ to $N=60$, more resonance peaks can be located in both panels of Fig.9(a) and Fig.9(b), showing that the increase of the length of the graphene segment between two normal metal sides can accommodates more quasi-bound states which are distributed at different angles.
When the strain strength is increased  to $\xi=0.2$ in Fig.9(b), more resonance peaks will emerge, which is consistent with the preceding analysis.
 
\begin{figure}[H]
\centering 
\subfigure[]{ 
\label{Fig10a} 
\includegraphics[height=4.5cm,width=0.7\textwidth]{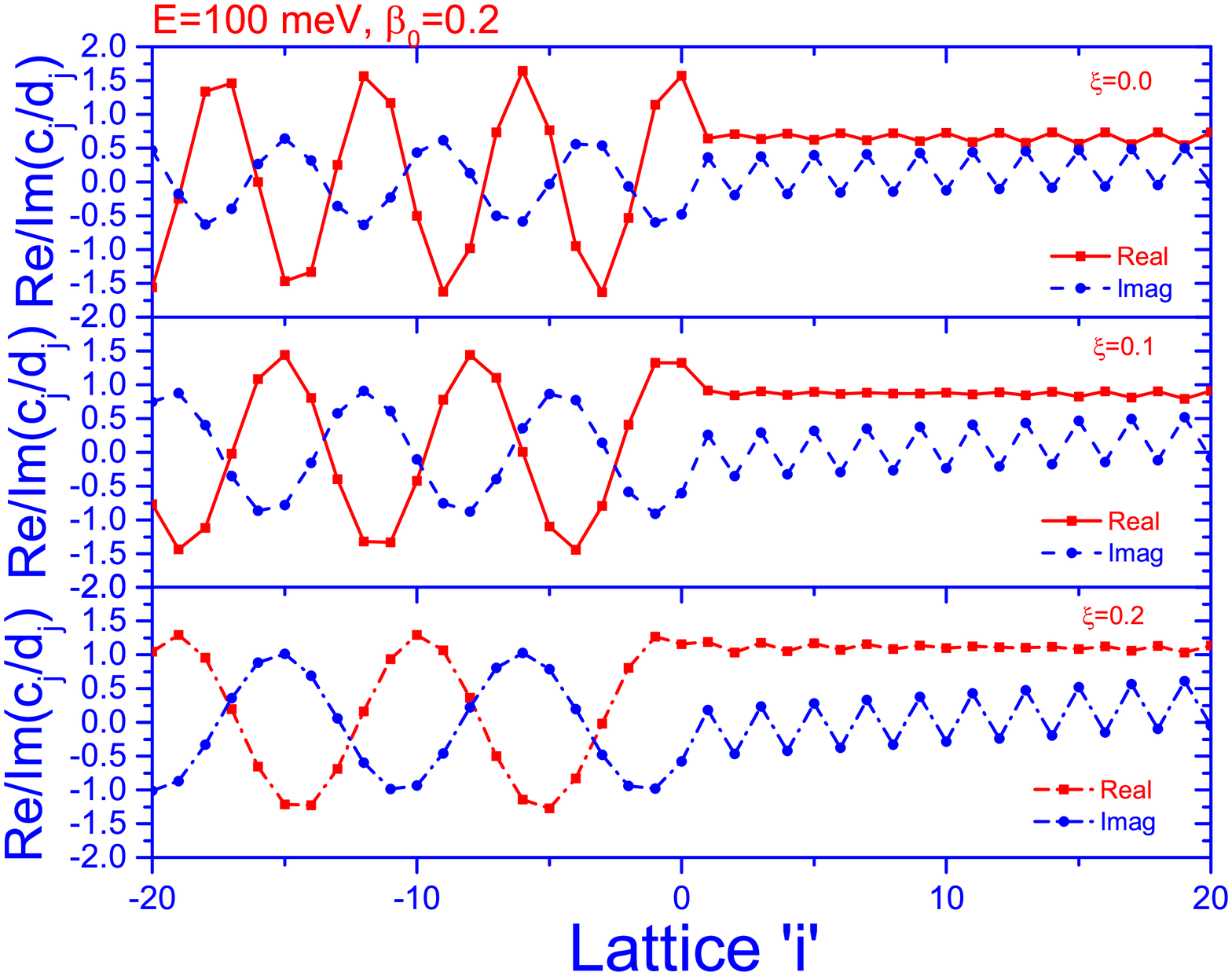}} 
\subfigure[]{ 
\label{Fig10b} 
\includegraphics[height=4.5cm,width=0.7\textwidth]{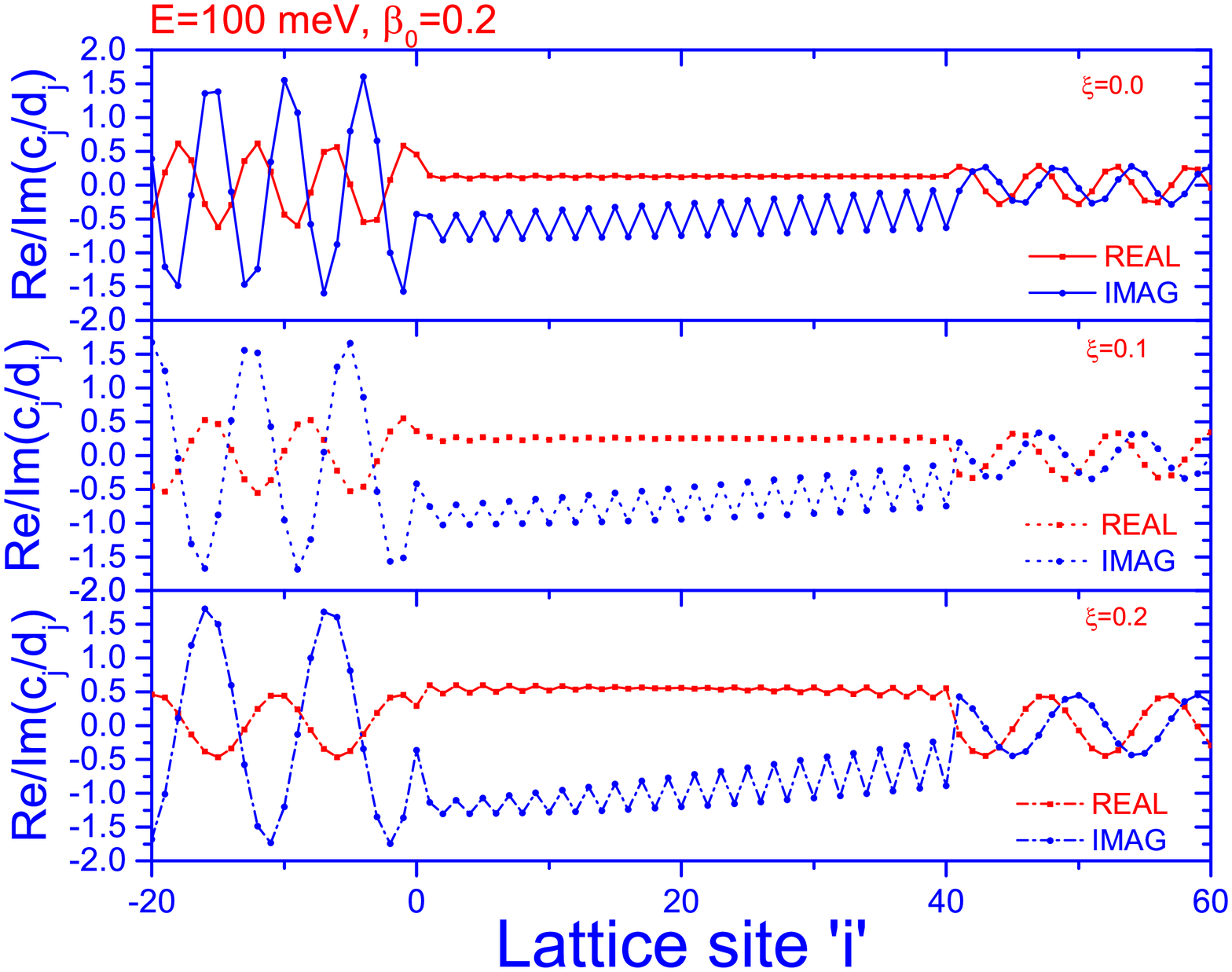}}
\caption{ Lattice wavefunctions for (a) NG junction and (b) NGN heterostructure under three different cases of strain $\xi=0.0,0.1,0.2$ with $E=100$ meV, $\beta_0=0.2$ and $q_x=q_y$, the line (red-solid) with square symbol stands for the real part of the wave function, while 
 the line (blue-dashed)  with dot symbol stand for the imaginary part of the wave function.  
}
\label{Fig.lable1} 
\end{figure}

The effective wavefunctions at the normal metal side and graphene side in the single NG junction can be derived with help of the following relation (for the meanings of the symbols, See Fig.2(a) for illustration), 
\begin{eqnarray}
\label{ver1_15}
\psi_0=\begin{pmatrix} d_{0}^{a} \cr d_{0}^{b} \end{pmatrix} =\begin{pmatrix} \frac{i 2\beta\lambda t_2\sin(k_s a_s)}{\beta t' \lambda^r_{ab}-Et'e^{-i(k_sa_s+q_x(a_g+u_1))}}\cr \frac{i 2\beta E\lambda^r_{ab} t_2\sin(k_s a_s)}{\beta t'(\lambda^r_{ab})^2-E\lambda^r_{ab} t'e^{-i(k_sa_s+q_x(a_g+u_1))}} \end{pmatrix}
\end{eqnarray} 
where Eq.(\ref{ver1_8}) and Eq.(\ref{ver1_9}) have been used (see Appendix for detail). The wavefunctions at the other lattices can be computed with 
 recurrence relationship defined in Eq.(\ref{ver1_9}), and the numerical results have been illustrated in three panels in Fig.10(a).
The incidence energy is chosen to be 100 meV, dimensionless $\beta_0$ is chose to be 0.2, and $q_x$ is set to be equal to $q_y$, the top, middle and bottom panels correspond to the cases of strain strength $\xi=0.0, 0.1, 02$ respectively. 
 On the metal side, the peaks of the real part of the wavefunction are almost coincident with the valleys of imaginary part of the wavefunction, and vice versa. Concerning with the amplitudes of the wavefunction on the graphene side, in contrast to the imaginary part of the graphene wavefunction, the real part of the graphene wavefunction has almost no trace of oscillations, but slightly increases due to the increase of the strain strength.  
Particularly, there is a remarkable phenomenon that the period of both the real and imaginary parts of the wavefunction on the metal side
 increases with the increases of the strain strength, indicating that the change of the period of the wavefunction on the metal side can serve as a sensor enabling the detection of the strain on the graphene side.

As for the NGN junction, on the left metal side, the relative phase between the real part and imaginary part is almost the same as that of the metal side in the single NG junction. Contrary to relative phase of the wavefunction on the left metal side, the relative phase difference between the real and imaginary parts of the wavefunction on the right metal side becomes smaller, and the oscillations amplitudes are almost identical to each other.
The periods of the wavefunction on either the left metal side and right metal side still becomes larger with the increase of the strain strength. 
 Such phenomenon can be interpreted by the longitudinal momentum match $k_s\cdot a_s=\cos^{-1}(-[E+2t_s\cos((K_y+q_y)a_s)]/2t_s)$ between the metal side and graphene side. To be specific, in the case of $q_y\ll K_y$ (for example, $q_y/K_y\approx 0.11$ at  incidence energy of $E=1000$ meV and $\xi=0.1$),  $(K_y+q_y)a_s$ is the  increasing function of the strain $\xi$, $\cos((K_y+q_y)a_s)$ is the decreasing function of the $\xi$, therefore $k_s$ is the decreasing function of $\xi$, which 
 can be deduced from the fact that $\cos^{-1}(-x)$ is an increasing function.  The decrease of $k_s$ with the increase of strain $\xi$ leads to the increase of the period  of the wavefunction on the metal side.

\section{Concluding remark}
  The transmission of the electron across the single NG and NGN junctions has been investigated, and the analytical expressions for the transmission of NG ang NGN junctions have been derived. The numerical analysis on the NG and NGN junctions have been carried out by focusing on the effects of strain strength, incidence energy, dimensionless hopping term on 
	 the transmission and conductance of the systems.
	 For the single NG junction, the profile of the maximum transmission  which has been extracted and plotted against the dimensionless interface hopping    respectively bears similarity to that of the conductance of the system which is also plotted against the dimensionless interface hopping. 
	The increase of the incidence energy shows negligible effect on the maximum transmission, but makes the maximum transmission angle subject to the wider range of the dimensionless interface hopping.
	The minor effect of the incidence energy on the transmission can also be found in the conductance of the single NG junction whose changing behavior poses a striking difference from that of the NGN junction. The reason behind such phenomenon can be attributed to the limited match 
 of the longitudinal ($Y$-direction) momentum between the two sides of the interface\cite{blanter} as well as lack of resonance mechanism which prevails at different angles in the graphene segment of NGN junction.

	As far as the NGN junction is concerned, the transmission shows more abundant structures when subjected to different incidence energies, dimensionless interface hoppings, and strain strengths. The increase of strain strength always induces more resonance peaks at different angles in transmission and can enhance the conductance. The increase of length of the middle graphene segment can accommodate more quasi-resonance states, leading to the more resonance peaks and richer structures in transmission.

		The real and imaginary parts of the wavefunction for both the single NG junction and NGN junctions have been computed under different strains, the relative phase and amplitudes of    wavefunction's real and imaginary parts at the metal sides and graphene side in both single NG and NGN junctions have been compared and  the relative phases between real and imaginary parts of the wavefunction on the two metal sides in the NGN junction show clearly different behavior with less difference in the outgoing wave on the right metal side. 
		The common phenomenon of particular interest in both single NG and NGN junctions is the increase of the wavefunction period on both metal sides due to the enhancement of the strain strength. That is to say,	the increase of the strain applied on the graphene side can make the transverse ($X$-direction) period of the wavefunction on the metal side increase, which can be interpreted by the variation of longitudinal ($Y$-direction) momentum due to strain.

	\appendix*

\section{Derivation of Eq.(\ref{ver1_7}),(\ref{ver1_8}),(\ref{ver1_9}),(\ref{ver1_15})}

\subsection{Derivation of Eq.(\ref{ver1_8})}
For a part of two-dimensional square lattice as shown in Fig.1(b), in the framework of the tight-binding approximation, 
the lattice wave function can be assumed to be
\begin{eqnarray}
|\underline{j},\underline{i}\rangle=|\underline{j}\rangle \exp(ik_y a_s \underline{i})
\end{eqnarray}  
where the conservation of the longitudinal momentum $k_y$ is always understood.
	Inserting the above relation into the single-particle version of ${\cal H}_s$ defined in section IIA\cite{gpzhang}, the one-dimensional effective Hamiltonian can be obtained as,
\begin{eqnarray}
~H^{\textrm{eff}}_{\textrm{NM}}=-t_s\sum\limits_{j}\left[\left(|j\rangle\langle j+1|+h.c.\right)+2\cos(k_ya_s)|j\rangle \langle j|\right]
\end{eqnarray}  
where the site energy $E_{j,i}$ has been assumed to vanish.
 Using expansion $\psi=\sum\limits_i c_i |i\rangle$ and Schr\"odinger equation $H\psi=E\psi$, as well as the orthonormality relation $\langle i|j\rangle=\delta_{ij}$, we can obtain 
\begin{eqnarray}
~-t_s \left[c_i\delta_{m,j}\delta_{i,m+1}+c_i\delta_{m,i}\delta_{j,m+1}+2\cos(k_y a_s)c_i \delta_{m,i}\delta_{j,m}\right]=E c_j
\end{eqnarray}  
where summation over repeated indices is employed.
Then Eq.(\ref{ver1_8}) can be derived as,
\begin{eqnarray}
\label{A4}
(E+2t_s \cos k_y a_s)c_j=-t_s(c_{j+1}+c_{j-1})
\end{eqnarray} 

\subsection{Derivation of Eq.(\ref{ver1_7}) and (\ref{ver1_9}) as well as boundary conditions for NG junction interface}
In order to derive Eq.(7) and (9), without loss of generality, assume that one $A$ atom is located at $('0',0)$
 and its nearest neighbors are $([1/2],\pm\frac{\sqrt{3}}{2}a_g)$  and $([-2/2],0)$; and one $B$ atom is located at $([4/2],0)$ and its nearest neighbors 
are $('3/2',\pm\frac{\sqrt{3}}{2}a_g)$ and $('6/2',0)$, and X-coordinates for A-atoms and B-atoms are respectively represented by index of single quotes $('i')$ and square bracket $[i]$ instead of the real X-cordinates as illustrated by the right panel of Fig.1(a).
For the cluster of atoms A and its three nearest neighbors, the effective one-dimensional Hamiltonian can be written as,
\begin{eqnarray}
~&& -2t_1 \cos\left(\frac{\sqrt{3}}{2}k_ya_g\right)\left(|'0'\rangle \langle [1/2]|+h.c. \right)- t_2 \left(|'0'\rangle \langle [-2/2]|+h.c.\right)
\end{eqnarray}  
where the relation $|'i',\underline{j}\rangle =|'i'\rangle \exp(ik_y \underline{j}a_g)$ has been used.
For the same reasoning, the effective Hamiltonian for the cluster of atom $B$ and its three nearest neighbors is,
\begin{eqnarray}
~-2t_1 \cos\left(\frac{\sqrt{3}}{2}k_ya_g\right)\left(|[4/2]\rangle \langle '3/2'|+h.c.\right)- t_2 \left(|[4/2]\rangle \langle '6/2'|+h.c.\right)
\end{eqnarray}  
By combining the results for $A$-atom centered cluster and $B$-atom centered cluster, it is found that 
 A-sublattice $L_a^i$ (indicated by blue dashed lines) and B-sublattice (indicated by red solid lines) $L_b^j$ in Fig.1(b) are of mutually Hermitian conjugate, that is to say, the same following effective Hamiltonian can be obtained by working on sublattice $L_a^i$ or $L_b^j$,
\begin{eqnarray}
~ H^{\textrm{eff}}_{\textrm{SG}}=&&-\sum\limits_i \bigg[\left(2t_1 \cos\left(\frac{\sqrt{3}}{2}k_ya_g\right) |'3i/2'\rangle \langle [3i/2+1/2]| + t_2 |'3i/2'\rangle \langle [3i/2-1]|\right) \nonumber \\ &&  +h.c.\bigg]
\end{eqnarray} 
Substitute the expansion $\psi=\sum\limits_j d_j^a |'j'\rangle + \sum\limits_m d_m^b|[m]\rangle$ into Schr\"odinger equation $E\psi=H^{\textrm{eff}}_{\textrm{SG}} \psi$, we can obtain 
\begin{eqnarray}
~&& -2t_1\cos\left(\frac{\sqrt{3}}{2}k_ya_g\right)\sum\limits_i\left[d^a_{3i/2}|[3i/2+1/2]\rangle+d^b_{3i/2+1/2}|'3i/2'\rangle\right] \nonumber \\ &&
-t_2 \left[d^a_{3i/2}|[3i/2-1]\rangle+d^b_{3i/2-1}|'3i/2'\rangle\right]=E \sum\limits_j d_j^a |'j'\rangle + \sum\limits_m d_m^b|[m]\rangle
\end{eqnarray}  
Using bra $\langle '3i/2' |$ and $\langle [3i/2+1/2]|$ to respectively multiply the two sides of the above equation,
then Eq.(\ref{ver1_7}) can be derived as,
\begin{eqnarray}\
\label{A9}
~&& Ed_{3i/2}^{a}=-t_2 d_{3i/2-1}^{b}-2t_1\cos\frac{\sqrt{3}k_ya_g}{2} d_{3i/2+1/2}^b    \nonumber \\ &&
Ed_{3i/2+1/2}^{b}=-t_2 d_{3i/2+3/2}^{a}-2t_1\cos\frac{\sqrt{3}k_ya_g}{2} d_{3i/2}^a
\end{eqnarray}
where the consistence with the graphical illustration of Fig.1(c) can be clearly seen.  
The derivation of the first two equations in Eq.(\ref{ver1_9}) can be accomplished by utilizing Fig2(a) and Eq.(\ref{A4}) where the index $j=0$ has been chosen,
\begin{eqnarray}
~(E+2t_s \cos k_y a_s)c_0=-t_s(c_{1}+c_{-1})
\end{eqnarray}  
Since the position of $c_1$ has been replaced by that of $d_0^a$ at the NG interface (see Fig.2(a)), then the first equation of the Eq.(\ref{ver1_9}) is derived as,
\begin{eqnarray}
~Ec_0=-2t_s \cos(k_y a_s)c_0-t_sc_{-1}-t'd_0^a
\end{eqnarray}  
where the original $t_s$ for term $c_1$ has further been replaced by the interface hopping $t'$ for the term $d_0^a$, leading to the consistence with the illustration of Fig2(a).
By setting $i=0$ in the first equation of Eq.(\ref{A9}),
\begin{eqnarray}
~Ed_0^a=-t_2 d_{-1}^{b}-2t_1\cos\frac{\sqrt{3}k_ya_g}{2} d_{1/2}^b
\end{eqnarray}  
Since the term $d_{-1}^b$ has been replaced by term $c_0$, the above equation can be rewritten as, 
\begin{eqnarray}
~Ed_0^a=-t' c_0-2t_1\cos\frac{\sqrt{3}k_ya_g}{2} d_{1/2}^b
\end{eqnarray}  
where the original $t_2$ for term $d^b_{-1}$ has also been replaced by the interface tunneling $t'$ for the term $c_0$.
The two boundary conditions can be derived by Eq.(A.10), (A.11) as well as by Eq.(A.12) and (A.13) respectively.
\begin{eqnarray}
~t'd_0^a=t_sc_1 ~~~~~~~~~~~~~~~~~t'c_0=t_2d^b_{-1}
\end{eqnarray}  

\subsection{Derivation of Eq.(\ref{ver1_15})}
For right-moving wave, substituting $d^b_{-1}=d_0^be^{-iq_x(a_g+u_1)}$ and $d^b_{1/2}=d_0^b e^{iq_x(\frac{a_g}{2}+u_2)}$ into Eq.(A.12), then the relation between $d_0^a$ and $d_0^b$ can be obtained as, 
\begin{eqnarray}
~Ed_{0}^{a} =\lambda^r_{ab} d_{0}^{b} 
\end{eqnarray}  
	where $\lambda^r_{ab} \equiv  -\left[t_2 e^{-iq_{x} \left(a_g +u_1 \right)} +2\left(t_1 \cos \frac{\sqrt{3} k_y a_g }{2} e^{iq_{x} \left(\frac{a_g }{2} +u_2 \right)} \right)\right]$.
		Substituting the wavefunction $c(x)=\left(e^{ik_s x}+r e^{-ik_s x}\right)$ $(x=0, x=a_s)$ on the metal side into the boundary conditions Eq.(A.14), then we can obtain the following equations,
\begin{eqnarray}
&& t_{s} \left(e^{ik_{s} a_s } +re^{-ik_{s} a_s } \right)=t'd_{0}^{a}  \nonumber \\ &&
t'\left(1+r\right)=t_2 e^{^{-iq_{x} \left(a_g +u_1 \right)} } d_{0}^{b},
\end{eqnarray} 
	Solving Eq.(A.15) and Eq.(A.16),  Eq.(14) can finally be obtained.

\subsection{The Low energy approximate expression for Eq.(\ref{ngexact}) }
		In the low energy regime ($q_x a_g, q_y a_g\ll 1$), all terms involving in $q_x,q_y$ can be expanded in the first order of $q_x,q_y$ $({\cal {O}}(q^2_x,q^2_y))$.
	Eq.(\ref{ngexact}) and $\lambda^r_{ab}$ can be approximately obtained as, 
\begin{eqnarray}
&& e^{-iq_{x} \left(a_g +u_1 \right)} =1-iq_{x} \left(a_g +u_1 \right) \nonumber \\ &&
e^{iq_{x} \left(\frac{a_g }{2} +u_2 \right)} =1+iq_{x} \left(\frac{a_g }{2} +u_2 \right)  \nonumber \\
&& \lambda^r_{ab} \approx
-t_g\left[-i \tilde{t}_2 q_{x} \left(\frac{3}{2} a_g +u_1+u_2 \right)
  -\frac{\sqrt{3} }{2} q_{y} a_g \sqrt{4\tilde{t}^2_1-\tilde{t}^2_2}\right] 
	\equiv t_g {\cal K}_1(q_x,q_y) 
\end{eqnarray}
where the dimensionless hopping terms $\tilde{t}_i\equiv t_i/t_g, (i=1,2)$.
Then substituting the above equation into the Eq.(\ref{ngexact}),  we can obtain the approximate reflectance as in the low energy regime,
\begin{eqnarray}
R=\left|r\right|^2 \approx  \left|\frac{2\beta{\cal K}_1(q_x,q_y)+e^{i k_{s} a_s}\sqrt{3}  a_g q\sqrt{4\sin^2\theta \left(\tilde{t}_1^2 -\tilde{t}_2^2 \right)+3\tilde{t}_2^2  } }{2\beta{\cal K}_1(q_x,q_y)+e^{-i k_{s} a_s }\sqrt{3}  a_g q\sqrt{4\sin^2\theta \left(\tilde{t}_1^2 -\tilde{t}_2^2 \right)+3\tilde{t}_2^2  }  } \right|^2 
\end{eqnarray}

\end{document}